\documentclass[
    aps,
    prd,
    twocolumn,
    superscriptaddress,
    nofootinbib,
    longbibliography,
    floatfix,
    preprintnumbers
]{revtex4-2}

\usepackage[utf8]{inputenc}
\usepackage[dvipsnames]{xcolor}
\usepackage{graphicx}
\usepackage{amsmath,amssymb,amsfonts,amsthm}
\usepackage{mathtools}
\usepackage{bm}
\usepackage{bbm}
\usepackage{microtype}
\usepackage{booktabs}
\usepackage{hyperref}
\definecolor{wine}{RGB}{136,34,85}
\definecolor{teal}{RGB}{0,85,102}

\hypersetup{
    colorlinks = true,
    citecolor = wine,
    linkcolor = teal,
    urlcolor = teal,
}

\newcommand{\be}{\begin{equation}}
\newcommand{\ee}{\end{equation}}
\newcommand{\bal}{\begin{align}}
\newcommand{\eal}{\end{align}}

\begin{document}

\preprint{IFT-UAM/CSIC-26-95}

\title{Search for Planetary-mass Black Holes with an Improved Viterbi Algorithm}

\author{Ra\'ul Rodr\'iguez}
\email{raul.rodriguez-dominguez@uib.cat}
\affiliation{IAC3, Universitat de les Illes Balears, Crta.Valldemossa km 7.5, E-07122 Palma, Spain}
\affiliation{Instituto de F\'isica Te\'orica (IFT) UAM-CSIC, C/ Nicol\'as Cabrera 13-15, Campus de Cantoblanco UAM, 28049 Madrid, Spain}

\author{George Alestas}
\email{g.alestas@csic.es}
\affiliation{Instituto de F\'isica Te\'orica (IFT) UAM-CSIC, C/ Nicol\'as Cabrera 13-15, Campus de Cantoblanco UAM, 28049 Madrid, Spain}

\author{Sachiko Kuroyanagi}
\email{sachiko.kuroyanagi@csic.es}
\affiliation{Instituto de F\'isica Te\'orica (IFT) UAM-CSIC, C/ Nicol\'as Cabrera 13-15, Campus de Cantoblanco UAM, 28049 Madrid, Spain}
\affiliation{Department of Physics and Astrophysics, Nagoya University, Nagoya, 464-8602, Japan}

\author{Juan Garc\'ia-Bellido}
\email{juan.garciabellido@uam.es}
\affiliation{Instituto de F\'isica Te\'orica (IFT) UAM-CSIC, C/ Nicol\'as Cabrera 13-15, Campus de Cantoblanco UAM, 28049 Madrid, Spain}

\date{\today}

\begin{abstract}
Primordial black holes in the planetary-mass range have attracted renewed interest; however, the search for gravitational waves from such binaries remains challenging due to their long-lived nature. In this work, we present, define, and validate a fully operational search pipeline developed to detect planetary-mass binaries during their inspiral phase. We use the Viterbi algorithm, a dynamic programming technique that recovers the most likely track based on a Hidden Markov Model. To enhance its performance, we introduce a novel time-frequency representation of the data and a candidate isolation procedure that separates signals from background noise. The evaluation of candidates is carried out using the two detection statistics, $n_{\sigma}$ and NMSE, which quantify the power significance and the consistency with the expected binary evolution. We then validate the search method using O3 LIGO Hanford data with a population of injected signals. The pipeline is able to recover most of the signals with a fixed false-alarm ratio of $3\%$, covering Galactic scales across most of the parameter space and reaching luminosity distances $ \gtrsim 100$ kpc in the most sensitive region. For each candidate, we also obtain an estimate of the system's chirp mass, whose accuracy remains high throughout the detectable range, enabling a rapid characterization of the system upon detection.
\end{abstract}

\maketitle
%
%--------------------------------------------------------------------------------------------------------------------------------------------------------------------------------------

\section{Introduction \label{sec:introduction}}
A new era in the study of black holes emerged with the discovery of gravitational waves (GWs)~\cite{LIGOScientific:2016aoc}. In particular, the routine detection of binary black hole mergers~\cite{LIGOScientific:2018mvr, LIGOScientific:2020ibl, PhysRevD.109.022001, PhysRevX.13.041039, Abac_2026,  LIGOScientific:2026wfs} is helping to characterize these bodies in a comprehensive way. This is one of the reasons for the renewed interest in the long-standing hypothesis of primordial black holes (PBHs). Such bodies would be formed in the early Universe, and would be able to explain some of the most intriguing questions of the current cosmological model~\cite{Hawking:1971ei, Carr:1974nx,Garcia-Bellido:1996mdl}. One of the most interesting aspects is that a specific mass distribution of PBHs could account for all or a fraction of the dark matter (DM)~\cite{Carr:2016drx, green2024primordialblackholesdark}. The emphasis on a mass distribution is important, as different PBH formation scenarios predict a wide variety of mass spectra spanning many orders of magnitude~\cite{Carr:2020gox}.

With the growing interest in PBHs~\cite{Carr:2023tpt}, and the hundreds of binary black hole mergers  detected so far~\cite{LIGOScientific:2026wfs}, GW observations have emerged as one of the most promising avenues for probing this DM candidate~\cite{Clesse:2016vqa,Garcia-Bellido:2019vlf}. 
Specifically, the discovery of a sub-solar mass (SSM) black hole would be strong evidence of the existence of PBHs. 

Recent results by the LIGO-Virgo-KAGRA (LVK) collaboration, such as low-spin black holes~\cite{Miller_2020,LIGOScientific:2026wfs}, indications that challenge the lower and upper mass gaps~\cite{LIGOScientific:2020ibl, LIGOScientific:2021usb,Abbott_2023} or SSM candidates~\cite{Morras_2023,Prunier_2023}, have further motivated the search of PBHs. However, current state-of-the-art search methods are unable to fully explore the challenging mass range below sub-solar mass, where the long duration of inspiral signals makes matched-filter searches computationally prohibitive~\cite{Magee:2018opb}. An increasing number of templates with enormous durations limits matched filtering SSM searches to masses above $0.2 \, M_{\odot}$~\cite{theligoscientificcollaboration2026searchesbinarymergerssubsolar}. 

When the component masses are sufficiently small, the search can be performed using continuous-wave (CW) search methods. In essence, for such light systems, the GW signal during the inspiral phase becomes similar to a CW. Constraints in the asteroid-mass range have been reported using these CW based methods~\cite{Miller:2021knj,KAGRA:2022dwb, LIGOScientific:2026plm}; however, their applicability is limited by a maximum allowed spin-up of $\dot{f} < 10^{-8}\,\mathrm{Hz\,s^{-1}}$, which corresponds to chirp masses less than $\mathcal{O}(10^{-5})\,M_\odot$.

Recently, the intermediate mass window, from $10^{-4} M_{\odot}$ to $10^{-1} M_{\odot}$, has attracted attention due to some observational evidences. The Subaru HSC observations of M31 \\ and the five-year OGLE microlesing survey toward the Galactic bulge have identified ultra-short microlensing events consistent with planetary-mass PBHs~\cite{Niikura:2019kqi, sugiyama2026microlensingconstraintsprimordialblack}. Moreover, the NANOGrav collaboration found evidence of a nano-Hertz stochastic GW background using pulsar timing arrays~\cite{Agazie_2023}, which may be linked to scalar-induced GWs associated with PBHs in the planetary-to-subsolar mass range \cite{Dom_nech_2022, domenech2026unifiedoriginprimordialblack}. 

Note that although several microlensing surveys~\cite{Tisserand_2007,Wyrzykowski_2011} have placed stringent constraints on the abundance of PBHs in this mass range, these limits can be relaxed under alternative assumptions, such as extended mass distributions~\cite{Carr:2023tpt,Carr_2026}, PBH clustering~\cite{Garcia-Bellido:2017xfd,Gorton:2022fyb, Belotsky_2019}, or uncertainties in Galactic DM distribution~\cite{Hawkins:2025mlo}. Furthermore, for a given PBH abundance, the expected number of PBH binaries in the Milky Way depends on the binary formation scenario and remains subject to theoretical uncertainty. Therefore, searches for PBH binaries remain important, as they can provide independent evidence for the existence of PBHs and help constrain their binary formation mechanisms.

In this paper, we aim to investigate the uncovered mass range between sub-solar compact binary coalescence (CBC) and CW search methods, mainly $10^{-4} M_{\odot} \lesssim M_{\text{PBH}} \lesssim 10^{-1} M_{\odot}$. We propose here a fully operational detection method, based on the Viterbi algorithm. Our search is still a CW search technique in essence, but it is specifically tailored to detect long-duration transient signals, such as those produced by low-mass PBH inspirals.

The pioneering work exploring part of this mass range in GW data~\cite{Miller:2020kmv,Miller:2021knj, Miller:2024jpo, Miller:2024fpo, SajithMenon:2025vpl, Miller:2025ote, s551-7pch} was carried out using the \textit{Generalized Frequency-Hough} algorithm, a standard technique in CW searches. An alternative approach, the Band-Sampled Data COmpact Binary Inspiral (BSD-COBI) method~\cite{Andres-Carcasona:2023zny}, has also been proposed, based on heterodyning techniques to track the signal evolution. Other methods~\cite{Wang:2025fmh, Wang:2025lpj} were developed to look for mini extreme mass ratio inspirals, comprised of exotic compact objects, as PBHs.

In contrast, our study employs the Viterbi dynamic programming algorithm~\cite{Viterbi}, which infers the most probable sequence of hidden Markov states in a model-agnostic framework. The Viterbi algorithm has already been successfully applied in CWs~\cite{KAGRA:2022dwb, Suvorova:2016rdc, Bayley:2019bcb, Bayley:2020zfa, Bayley:2022hkz}, mainly through the \texttt{SOAP} package~\cite{Bayley:2019bcb}, a fully operational pipeline that extracts the most likely hidden signal from noisy GW  data.

The Viterbi algorithm is computationally efficient, as it operates in a model-agnostic framework without requiring a precomputed bank of templates. Due to its agnostic nature, the algorithm remains sensitive to signals with more complex waveform features, including higher-order corrections or additional physical effects such as dark-matter environments.

A recent work~\cite{Alestas:2024ubs} demonstrated the effectiveness of the Viterbi algorithm in searching for planetary-mass PBH inspirals assuming Gaussian noise. Here, we extend this approach by developing a fully operational search pipeline for this mass range. We introduce a new time-frequency representation of the data that improves performance and accuracy of the Viterbi algorithm. In addition, we develop a candidate-isolation procedure to discriminate potential signals from background noise. We also introduce new detection statistics that enhance the significance of the search and improve the discrimination between genuine signals and noise fluctuations. The resulting pipeline enables end-to-end analyses of real data, from the strain time series to candidate identification and evaluation.

The paper is structured as follows. In Sec.~\ref{sec:methodology}, we describe the key components of the search, including the properties of long-inspiral signals, the Viterbi framework adopted in this work, the adaptation of short-time Fourier transforms (SFTs) to this class of signals, and the detection statistics used to identify significant candidates. In Sec.~\ref{sec:Analysis}, we present the full pipeline architecture and evaluate its performance using O3b LIGO Hanford data~\cite{O3data}. Finally, in Sec.~\ref{sec:conclusions}, we summarize our results and discuss prospects for future searches in upcoming observing runs. The pipeline developed is publicly available at~\cite{viterbi_pbh_github}.

%--------------------------------------------------------------------------------------------------------------------------------------------------------------------------------------

\section{Methodology \label{sec:methodology}}
\subsection{Long Inspirals \label{sec:longinspirals}}
When two compact objects start inspiraling, orbital energy is radiated away via GW emission leading to an increase in the orbital angular frequency $\omega_{\rm orb}$, and hence in the GW frequency $f_{\rm gw}$. Assuming a circular orbit, we find the increase in frequency~\cite{Maggiore_Vol1}:
\begin{equation}
\dot{f}_{\text{gw}} = \frac{96}{5} \, \pi^{8/3} \left( \frac{G \mathcal{M}}{c^3} \right)^{5/3} f_{\text{gw}}^{11/3},
\label{dotfw}
\end{equation}
where $\mathcal{M}=\frac{(m_1 . m_2)^{3/5}}{(m_1 + m_2)^{1/5}}$ is the chirp mass of the system. For the most massive PBHs considered here, the $f_{\rm gw}$ at the inner most stable circular orbit (ISCO) lies well outside the detectable frequency band. As the binary is still far from merger, higher post-Newtonian order corrections are not required. This guarantees that the frequency evolution of Eq.~\eqref{dotfw} remains valid throughout the entire frequency band. Integrating Eq.~\eqref{dotfw}, we find the GW frequency evolution as a function of time:
\begin{equation}
 f_{\text{gw}} (t) = \frac{1}{\pi}\left( \frac{5}{256} \right)^{3/8} \left( \frac{G \mathcal{M}}{c^3} \right)^{-5/8} (t_{\text{coal}} - t)^{-3/8} .
\label{fgw(t)}
\end{equation}
We identify that $f_{\rm gw}$ formally diverges at a finite value of time, $t_{\rm coal}$, the \textit{coalescence time}. Then, strictly speaking, the frequency evolution is a function of the \textit{time to coalescence}, $(t_{\rm coal}-t)$, i.e., the remaining time until the merger. Equivalently, we can write
\begin{equation}
 (t_{\text{coal}} - t)  =\left( \frac{5}{256} \right) \left( \frac{G \mathcal{M}}{c^3} \right)^{-5/3}  (\pi f_{\text{gw}})^{-8/3}.
\label{ttocoal}
\end{equation}
\begin{figure}[h]
    \centering
    \includegraphics[width=\linewidth]{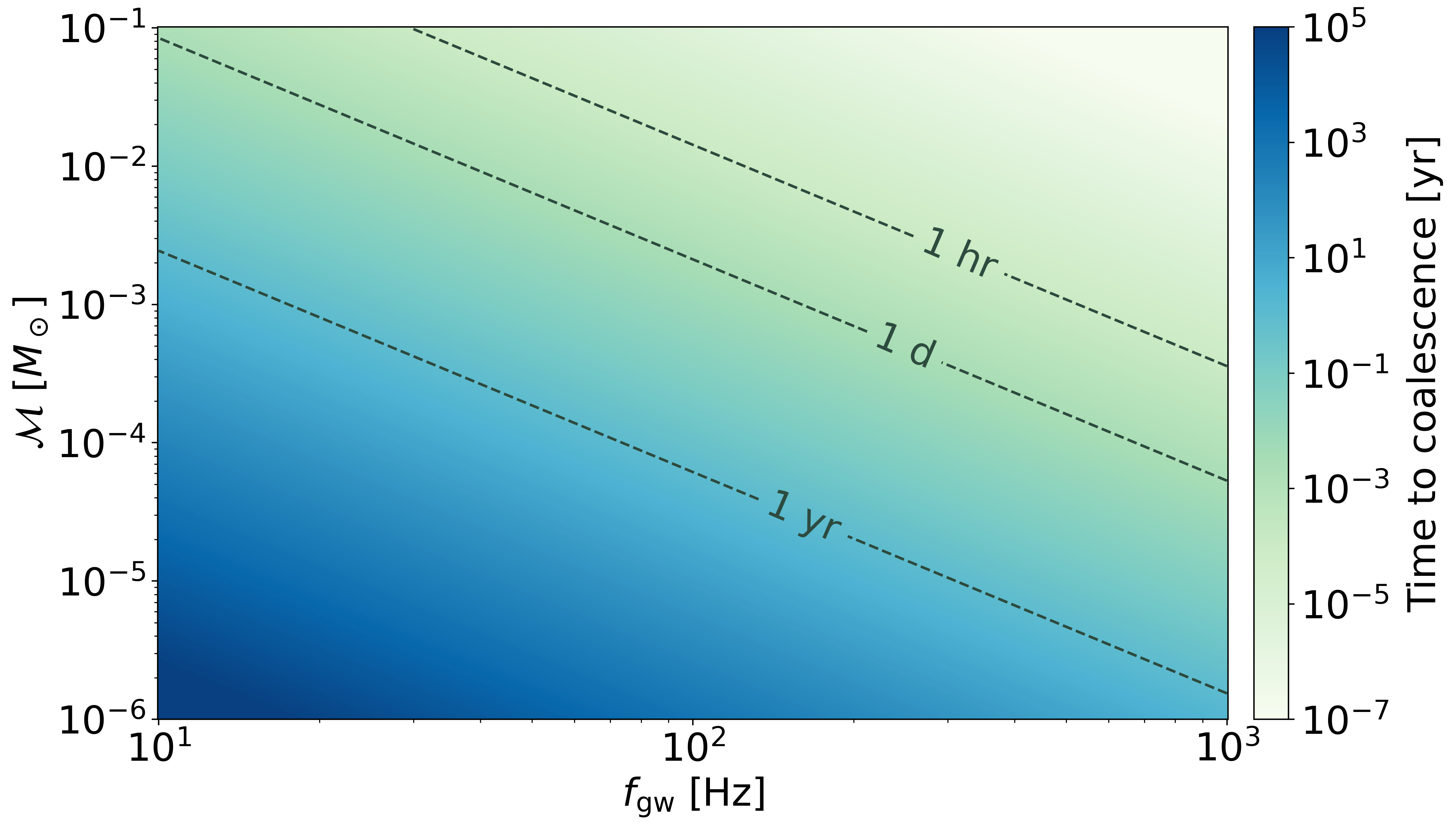}
    \caption{Time to coalescence as a function of the chirp mass and GW frequency of the binary system. The broad range of signal durations across the parameter space indicates that distinct techniques are required to effectively investigate the presence of PBHs at different mass scales.}
    \label{t_to_coal}
\end{figure}

Due to their small masses, planetary-mass PBHs are expected to undergo inspirals that last significantly longer in the detectable frequency band than typical BBHs detected by the LVK collaboration, potentially spanning months or even years~\cite{Miller:2020kmv}. This can be readily visualized by expressing the \textit{time to coalescence} as a function of the binary chirp mass and the GW frequency, as shown in Fig.~\ref{t_to_coal}. The figure clearly illustrates that PBH inspirals can remain in the detector band for extended periods, making them particularly well suited for CW and transient CW search methods.

\begin{figure}[h]
    \centering

    \includegraphics[width=\linewidth]
    {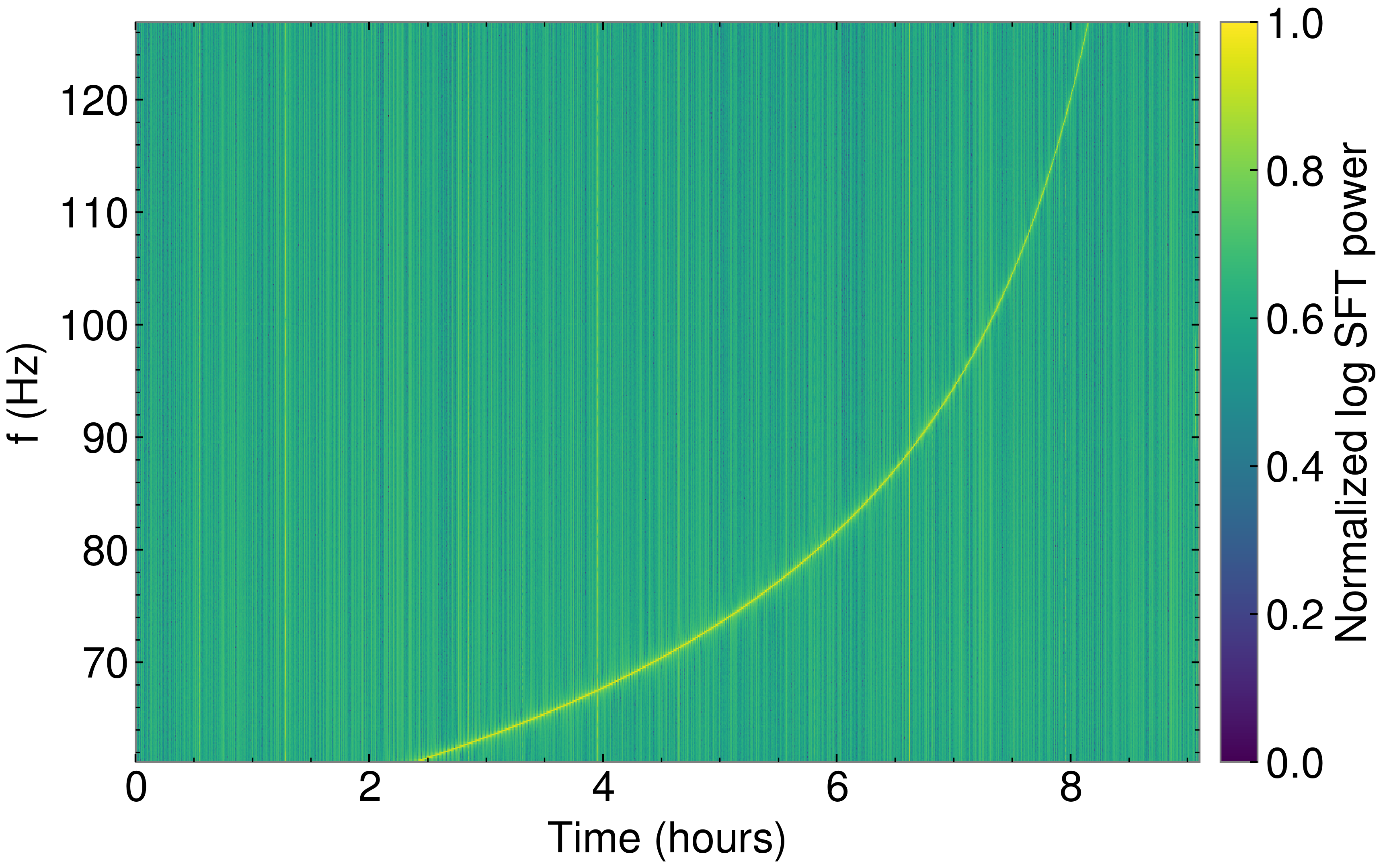}

    \vspace{1ex}

    \includegraphics[width=\linewidth]
    {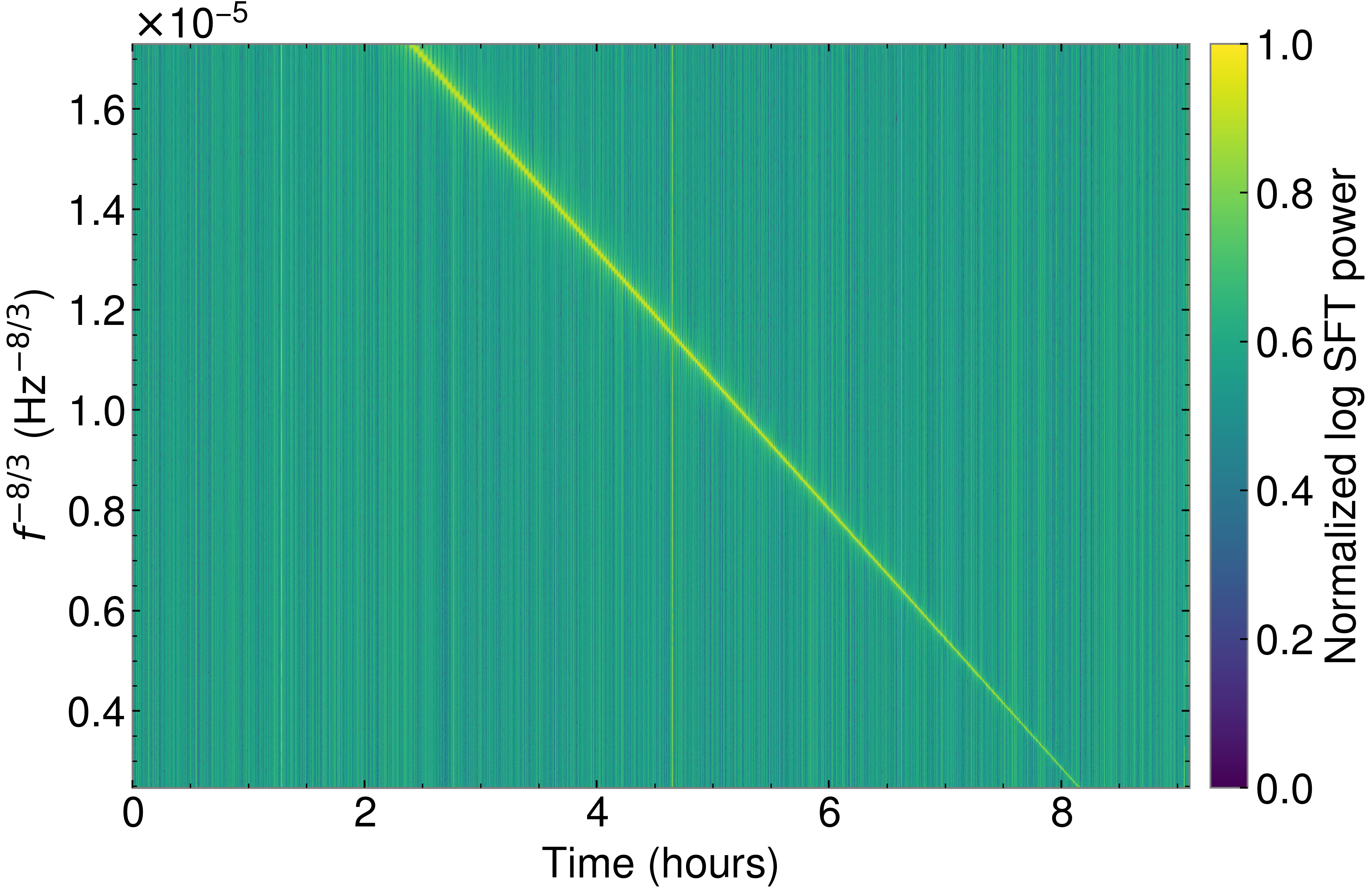}

    \caption{
    Time-frequency representation of O3b LIGO Hanford data with a high-SNR signal injected.
    Top panel: Spectrogram of a strain time series with a visible chirp of a long inspiral.
    Bottom panel: New frequency map ($f^{-8/3}$) of the strain, where the characteristic chirp becomes a straight line.
    }
    \label{fig:newmap}
\end{figure}

One of the novel aspects of this work is that we do not operate on conventional time--frequency maps. In standard spectrograms, an inspiral signal traces the characteristic chirp pattern shown in the top panel of Fig.~\ref{fig:newmap}, with a frequency evolution governed by Eq.~\eqref{dotfw}. This evolution can be rewritten as:
\begin{equation}
\frac{\dot{f}_{\text{gw}} }{f_{\text{gw}}^{11/3}}= C \cdot \mathcal{M}^{5/3},
\label{ctedotf}
\end{equation}
where $C$ is a constant value. We now introduce the coordinate transformation
\begin{equation}
 \frac{d}{dt} \left( f_{\text{gw}}^{-8/3} \right) = K \cdot \mathcal{M}^{5/3}, 
\label{dotfgwnew}
\end{equation}
where $K$ is the constant
\begin{equation}
K = - \frac{256}{5} \pi^{8/3} \left(\frac{G}{c^3}\right)^{5/3}.
\label{eq:cte}
\end{equation}

In Eq.~\eqref{dotfgwnew}, we just have one independent variable, the chirp mass. This naturally motivates the introduction of a new frequency map, $(t, f^{-8/3})$, where the long inspirals become straight lines as seen in the bottom panel of Fig.~\ref{fig:newmap}. Furthermore, in these transformed coordinates, 
the track gradient is directly determined by the chirp mass of the system. 
Both features play a key role in improving the performance of the search method.

\subsection{The Viterbi Algorithm \label{sec:viterbi}}
The Viterbi algorithm is a dynamic programming algorithm that allows to infer the most probable sequence of hidden states in a Markov model based on noisy data. It operates within the framework of Hidden Markov Models, where the true state of the system evolves according to a Markov process, but is not directly observable. In this work, we use the Viterbi algorithm as a novel technique for searching long-lasting inspirals, following the idea proposed in~\cite{Alestas:2024ubs}. The approach is implemented through the \texttt{SOAP} package~\cite{Bayley:2019bcb, Bayley:2020zfa, Bayley:2022hkz, soapcw}, which has been originally developed to search for CW signals.

In the \texttt{SOAP} implementation, the input frequency time-series data is divided into $N$ segments of equal duration~\cite{Bayley:2019bcb}, forming the dataset of time series segments $\mathbf{x}_i$, which we denote as $D \equiv \{\mathbf{x}_i\}$. Each index $i$ corresponds to a specific time segment. The track we aim to recover is the sequence of signal frequencies $\boldsymbol{\nu} \equiv \{\nu_i\}$, where $\nu_i$ represents the GW frequency within segment $\mathbf{x}_i$. The goal is to evaluate all possible frequency tracks and identify the one that maximizes the posterior probability, i.e.\ the most likely signal path given the data:
\begin{equation}
\hat{\boldsymbol{\nu}} = \arg \max_{\boldsymbol{\nu}} p(\boldsymbol{\nu} \mid D).
\end{equation}
Following Bayes' theorem, the posterior probability takes the form:
\begin{equation}
p(\boldsymbol{\nu} \mid D) = \frac{p(\boldsymbol{\nu})\, p(D \mid \boldsymbol{\nu})}{p(D)},
\label{eq:posterior}
\end{equation}
where $p(\boldsymbol{\nu})$ is the prior probability of the track,  $p(D \mid \boldsymbol{\nu})$ is the observation likelihood (the probability of the data given the track) and $p(D)$ is the model evidence. Notice that $p(D)$ is constant for all $\boldsymbol{\nu}$, so maximizing $ p(\boldsymbol{\nu} \mid D)$ is equivalent to just maximizing the numerator, i.e. the joint probability $ p(\boldsymbol{\nu}, D)$.

We can expand explicitly the joint probability and apply the observational independence and the Markov property~\cite{article}, leading to 
\begin{equation}
p(\boldsymbol{\nu}, D)=p(\nu_0)\prod_{n=1}^{N} p(\nu_n \mid \nu_{n-1})\prod_{n=0}^{N} p(x_n \mid \nu_n),
\end{equation}
where $p(\nu_n \mid \nu_{n-1})$ is the ``transition'' probability for $\nu_n$ given the frequency at the last step is $\nu_{n-1}$. Now, for numerical stability, we work in the logarithmic domain. The most probable signal track $\hat{\boldsymbol{\nu}}$ that maximizes the posterior probability, the \textit{Viterbi track}, is then found by maximizing the logarithm:
\begin{align}
\hat{\boldsymbol{\nu}} &= \arg\max_{\boldsymbol{\nu}} \left[ \log p(\boldsymbol{\nu}, D) \right] 
 \\ &\quad =\arg\max_{\boldsymbol{\nu}} \Bigg\{ 
\log p(\nu_0) + \log p(\mathbf{x}_0 \mid \nu_0) \notag 
 + \\ &\quad \quad \sum_{i=1}^{N} \left[ 
\log p(\nu_i \mid \nu_{i-1}) + \log p(\mathbf{x}_i \mid \nu_i) 
\right] \Bigg\}.
\end{align}
Rather than evaluating every path, it recursively computes the maximum probability of arriving at each state at each time step, keeping track of the most likely predecessor, drastically reducing the search space. In fact, for a state space of size $S$ and a sequence of length $N$, a naive search would require evaluating $\mathcal{O}$($S^N$) possibilities, while Viterbi achieves the same result with $\mathcal{O}$($S^2N$) operations~\cite{article}.

The transition matrix $T$ encodes the prior log-probabilities $\log p(\nu_n \mid \nu_{n-1})$, which represent the probability of the signal's frequency transitioning from the $(n-1)$th to the $(n)$th state. It provides a natural framework for imposing loose model constraints. For instance, some recent works~\cite{Bayley:2019bcb, Alestas:2024ubs} restricted successfully the track to move at most one bin per step: up, center, or down (UCD) transition or ``jump'', reducing $T$ to three distinct values. It is a $3\times1$ matrix, where each row represents the probability of the next jump. This configuration was chosen based on the expected smoothness of the signal. As we will see later, we will take advantage of the transition matrix to improve Viterbi's performance.

\subsection{Short-time Fourier Transform  \label{sec:sft}}
The primary input to the Viterbi algorithm is the set of Short-Time Fourier Transforms (SFTs) computed from the strain time-series data frames. These are obtained by dividing the full data stream $s$[n], where $n$ indexes discrete time samples, into segments of duration $T_{\mathrm{\rm SFT}}$, applying a window function $w[n]$, and performing a Discrete Fourier Transform on each segment. This process yields a time-frequency representation $\tilde{s}[n,k]$, where $k$ labels frequency bins. The frequency resolution is directly defined as $\Delta f = 1 / T_{\rm SFT}$, while the total number of SFTs is given by $N_{\rm SFT} = T_{\rm obs}/T_{\rm SFT}$, where $T_{\rm obs}$ the total observation time. 

In the context of CW and transient-CW search techniques, it is typically assumed that the signal frequency remains confined within a single frequency bin during each SFT. For a given SFT duration and an inspiral signal from a binary system with chirp mass $\mathcal{M}$, the maximum frequency $f_{*}$ that satisfies the condition~\cite{Alestas:2024ubs} that the signal remains within the same frequency bin is given by
\begin{equation}
f_* = \frac{1}{\pi} \left( \frac{5\pi}{96} \right)^{3/11} \left( \frac{G \mathcal{M}}{c^3} \right)^{-5/11} T_{\mathrm{\rm SFT}}^{-6/11} \,.
\label{criticalfreq}
\end{equation}
Above this frequency, the frequency evolution becomes sufficiently rapid as the system approaches the merger phase, causing the signal to drift out of a frequency bin.

Eq.~\eqref{criticalfreq} indicates that shorter SFT durations allow access to higher maximum frequencies, enabling the inclusion of more rapidly evolving signals and thus potentially increasing the sensitivity. On the other hand, reducing the SFT duration increases the number of segments, and hence the number of degrees of freedom. As a consequence, the statistical fluctuations of the detection statistic become larger, reducing its statistical significance. These two competing effects imply the existence of an optimal choice of SFT duration (or equivalently, an optimal maximum frequency) that balances sensitivity to high-frequency signals and statistical stability.

The optimal SFT duration was derived in~\cite{Alestas:2024ubs} by maximizing the detection horizon distance. It was shown~\cite{Alestas:2024ubs}  that the optimal frequency range can be obtained by maximizing the following function:
\begin{equation}
    F(f_0, f_*) = \frac{f_0^{2/3}}{f_*^{11/24}} \sqrt{\int_{f_0}^{f_*} \frac{df}{f^{7/3} S_n (f)}}.
    \label{eq:F_f0fs}
\end{equation}
Here, $f_0$ and $f_*$ denote the low- and high-frequency cutoffs, respectively. Note that, by construction, the shape of Eq.~\eqref{eq:F_f0fs} is similar to the signal-to-noise ratio (SNR). This function is determined solely by the shape of the noise power spectral density (PSD) and does not depend on the signal parameters, such as the chirp mass. For the LIGO Hanford O3 PSD~\cite{O3data, abbott2020prospects}, we find that the function is maximized at $f_{0, \mathrm{opt}} = 61.1\,\mathrm{Hz}$ and $f_{*, \mathrm{opt}} = 126.8\,\mathrm{Hz}$ \footnote{For the LIGO O5 PSD the optimum frequency band lies in a similar region, with  $f_{0, \mathrm{opt}} = 57.5\,\mathrm{Hz}$ and $f_{*, \mathrm{opt}} = 134.4\,\mathrm{Hz}$.}.

Using Eq.~\eqref{criticalfreq} and normalizing it with respect to the optimal high-frequency cutoff $f_{*,\mathrm{opt}}$, we obtain the optimal SFT duration for the O3 dataset:
\begin{align}
    T_\mathrm{\rm SFT}^\mathrm{opt} & =  \left(\frac{5 \pi}{96}\right)^{1/2} \left(\frac{G \mathcal{M}}{c^3} \right)^{-5/6} (\pi f_{*, \mathrm{opt}})^{-11/6} 
    \nonumber \\  & = 8.50 ~ \mathrm{s} \left(\frac{\mathcal{M}}{10^{-2} M_\odot}\right)^{-5/6} \left(\frac{f_{*, \mathrm{opt}}}{126.8 ~\mathrm{Hz}}\right)^{-11/6}.
    \label{eq:TSFTopt}    
\end{align}
See Ref.~\cite{Alestas:2024ubs} for a more detailed derivation.

As the optimum SFT duration depends on the chirp mass under consideration, searching over a range of chirp masses requires the use of multiple values of $T_{\text{SFT}}$. Consequently, the optimal SFT duration becomes an additional unknown parameter of the search, being directly linked to the chirp mass of the source.
 
\subsection{Detection Statistics \label{sec:detstatistics}}
In order to identify significant candidates, we require a 
detection statistic capable of quantifying how unlikely the
recovered Viterbi track ($\hat{\boldsymbol{\nu}}$) is under the noise-only hypothesis.
Since the algorithm selects the most probable path in the
time--frequency map, the relevant quantity is not the power
accumulated along an arbitrary track, but the maximum
accumulated power among all the tracks explored by Viterbi.
Following a previous work ~\cite{Alestas:2024ubs}, we use the statistic $n_{\sigma}$ as one of the main ranking quantities of the search.

Let us consider a candidate track $\boldsymbol{\nu}=\{\nu_i\}$, where the index $i$ labels the SFT segment. For each point of the track, we define the matched-filter-like SNR contribution $\rho^{\rm mf}_i$, and the total incoherent SNR accumulated along the track as
\begin{equation}
    \rho_{\rm tot}^2(\boldsymbol{\nu})
    =
    \sum_{i=1}^{N_{\rm SFT}}
    \left|\rho^{\rm mf}_i(\nu_i)\right|^2 .
    \label{eq:rho_tot_track}
\end{equation}
The Viterbi algorithm returns the track that maximizes the
posterior probability, and therefore the relevant quantity for
candidate selection is
\begin{equation}
    \rho_{\rm tot,max}^2
    =
    \max_{\boldsymbol{\nu}}\,
    \rho_{\rm tot}^2(\boldsymbol{\nu}) .
    \label{eq:rho_tot_max}
\end{equation}
In the absence of a signal, $\rho_{\rm tot,max}^2$ does not
follow the same distribution as the accumulated power of a
single fixed track, since Viterbi maximizes over a large
number of correlated paths. This effect must be included in
the definition of the detection statistic.

We define $n_{\sigma}$ as the number of standard
deviations by which the recovered Viterbi SNR exceeds the expectation from a noise-only background,
\begin{equation}
    n_{\sigma}
    =
    \frac{
    \rho_{\rm tot,max}^2
    -
    \mu(\rho_{\rm tot}^{\rm opt}=0,N_{\rm SFT})
    }{
    \sigma(\rho_{\rm tot}^{\rm opt}=0,N_{\rm SFT})
    } .
    \label{eq:nsigma_definition}
\end{equation}
Here, $\mu(\rho_{\rm tot}^{\rm opt}=0,N_{\rm SFT})$ and
$\sigma(\rho_{\rm tot}^{\rm opt}=0,N_{\rm SFT})$ are,
respectively, the mean and standard deviation of the
noise-only distribution of $\rho_{\rm tot,max}^2$, for a
given number of SFTs. The condition
$\rho_{\rm tot}^{\rm opt}=0$ denotes the absence of an
injected signal. In practice, these quantities encode the
background expected from random noise fluctuations after
the Viterbi maximization.

This statistic provides an intuitive measure of the
significance of a recovered track. A candidate with large
$n_{\sigma}$ corresponds to a path whose accumulated power
is far from the typical noise-only maximum, and is therefore
less likely to be produced by a random fluctuation. This is
particularly useful in our search, since each choice of
$T_{\rm SFT}$ produces a different number of SFTs, and hence
a different noise background. The dependency on $N_{\rm SFT}$ in
Eq.~\eqref{eq:nsigma_definition} allows us to compare
candidates obtained from different time--frequency
representations.

\begin{figure}[h]
    \centering
    \includegraphics[width=\linewidth]{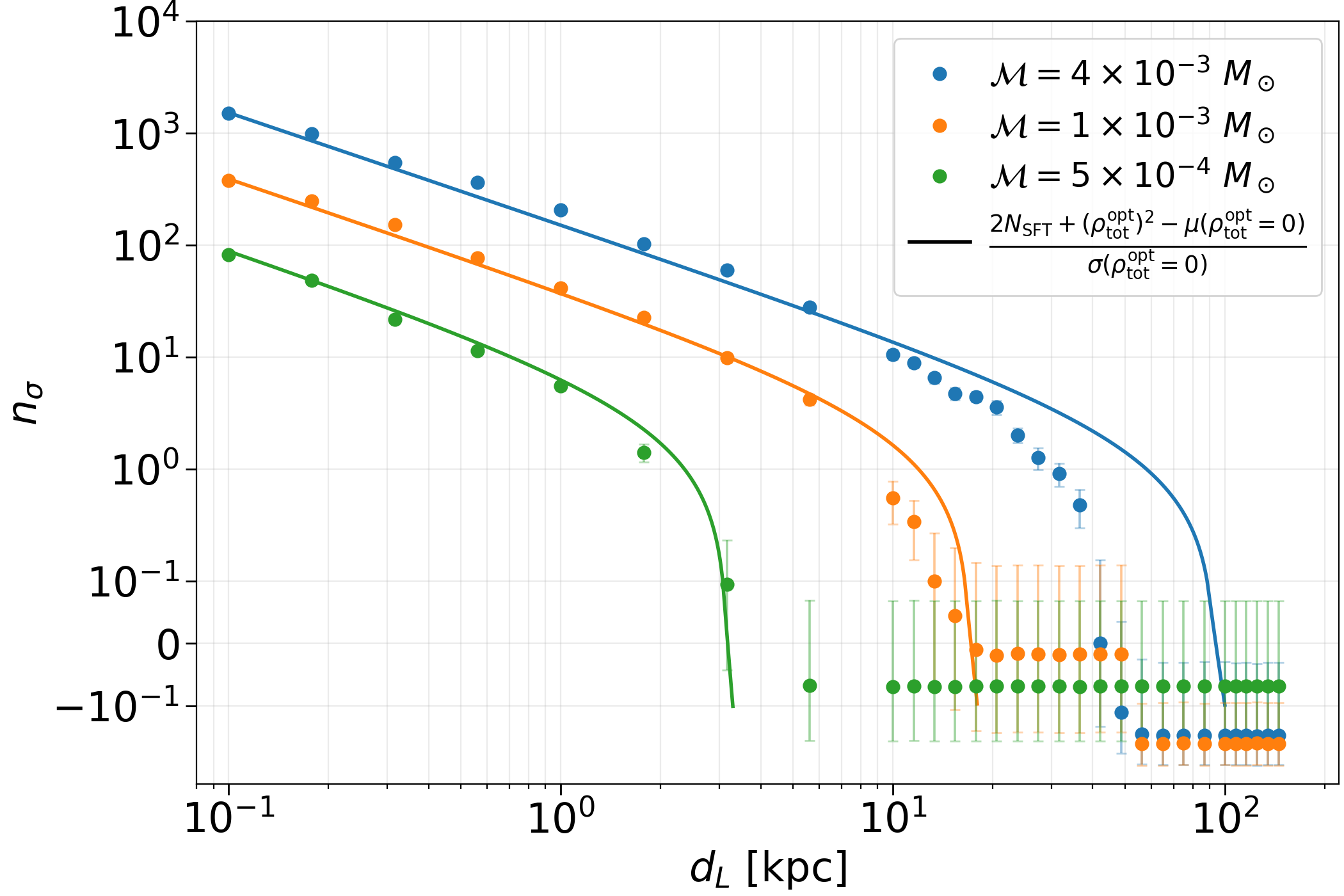}
    \caption{Evolution of the detection statistic $n_{\sigma}$ with luminosity distance. Points with error bars show the mean values recovered from 1000 hours of O3 LIGO Hanford data with injected signals, computed using Eq.~\eqref{eq:nsigma_definition}. The solid line corresponds to the analytical expected value of Eq.~\eqref{eq:nsigma_analytic}. As the luminosity distance increases, the injected signals become weaker, the loud signal assumption breaks down, and $n_{\sigma}$ approaches the noise-dominated regime.}
    \label{fig:nsigma_evolution}
\end{figure}

For a sufficiently loud signal, the Viterbi path is expected
to coincide with the true signal track. In this limit, the mean
accumulated SNR along the recovered track can be
approximated by the mean of the non-central
$\chi^2$ distribution~\cite{Alestas:2024ubs}, given by
\begin{equation}
    \left\langle \rho_{\rm tot,max}^2 \right\rangle
    \simeq
    2N_{\rm SFT}
    +
    \left(\rho_{\rm tot}^{\rm opt}\right)^2 ,
    \label{eq:rho_tot_loud_signal}
\end{equation}
where $\rho_{\rm tot}^{\rm opt}$ is the optimal total SNR of
the signal. Substituting this expression into
Eq.~\eqref{eq:nsigma_definition}, we obtain the analytical
approximation
\begin{equation}
    \left\langle n_{\sigma} \right\rangle
    \simeq
    \frac{
    2N_{\rm SFT}
    +
    \left(\rho_{\rm tot}^{\rm opt}\right)^2
    -
    \mu(\rho_{\rm tot}^{\rm opt}=0,N_{\rm SFT})
    }{
    \sigma(\rho_{\rm tot}^{\rm opt}=0,N_{\rm SFT})
    } .
    \label{eq:nsigma_analytic}
\end{equation}
This expression corresponds to the large-SNR limit in which
the recovered Viterbi track follows the physical signal. As
the signal becomes fainter, the loud signal assumption breaks down, and so does this approximation, since the algorithm can instead lock onto a noise-dominated path. In Fig.~\ref{fig:nsigma_evolution}, we show the evolution of this metric as a function of the luminosity distance, 
comparing the values recovered from data with injected signals (dots) and the analytical approximation (solid lines). One can detect how the approximation breaks once the signal enters the low-SNR regime. Nevertheless, Eq.~\eqref{eq:nsigma_analytic} provides the theoretical basis for interpreting $n_{\sigma}$ as a significance-like statistic and motivates its use as one of the main quantities for ranking candidates in this work.

In addition to the power-based statistic $n_{\sigma}$, we
introduce a second statistic aimed at quantifying the
morphological consistency of the recovered track with the
expected frequency evolution of a compact-binary inspiral.
This is particularly important in the low-SNR regime,
where the accumulated power along the Viterbi path can be
comparable to ordinary noise fluctuations. In this case,
$n_{\sigma}$ alone may not be sufficient to distinguish a real
slowly evolving signal from a noise-induced track.

For this purpose, we use the normalized mean square error (NMSE). Let
\begin{equation}
    \hat{\boldsymbol{y}}
    =
    \{\hat{y}_i\}_{i=1}^{N_{\rm seg}}
\end{equation}
be the portion of the Viterbi track selected as a candidate,
written in the remapped frequency coordinate $y \equiv f^{-8/3}.$ Here, $N_{\rm seg}$ denotes the number of time bins in the
candidate segment. From Eq.~\eqref{dotfgwnew},
the expected inspiral evolution in this coordinate is
approximately linear,
\begin{equation}
    y_{\rm model}(t, \mathcal{M})
    =
    K \cdot  \mathcal{M}^{5/3}(t-t_0),
    \label{eq:nmse_model}
\end{equation}
where $K$ is the constant introduced in Eq.~\eqref{eq:cte}, $\mathcal{M}$ is the chirp mass, and $t_0$ is the initial time of the candidate track. Therefore, in the transformed map, fitting the candidate morphology reduces to testing whether the recovered Viterbi track is consistent with a ``straight line'' whose slope is determined by the chirp mass. Strictly speaking, this assumption is not exact, as Doppler modulations introduce deviations from a purely linear frequency evolution. However, given the short $T_{\rm SFT}$ values considered in this work, these deviations are negligible and the signal track can be approximated as linear.

For a given value of the chirp mass, we define the NMSE as
\begin{equation}
    {\rm NMSE}
    =
    \frac{
    \sum_{i=1}^{N_{\rm seg}}
    \left[
    \hat{y}_i
    -
    y_{\rm model}(t_i,\mathcal{M})
    \right]^2
    }{
    \sum_{i=1}^{N_{\rm seg}}
    \hat{y}_i^{\,2}
    } .
    \label{eq:nmse_definition}
\end{equation}
The normalization by the total squared amplitude of the
recovered track makes the statistic dimensionless and
allows us to compare candidates located in different
frequency regions. Small NMSE values indicate that the
track follows the expected inspiral morphology, whereas
large NMSE values correspond to tracks that are poorly
described by the compact-binary frequency evolution.

Since the chirp mass of the source is not known a priori,
we minimize the NMSE over the target mass range,
\begin{equation}
    {\rm NMSE}_{\rm min} \equiv \min_\mathcal{M} \, {\rm NMSE (\mathcal{M})} ,     \label{eq:nmse_min}
\end{equation}
with
\begin{equation}
     \mathcal{M}\in [10^{-4}, 10^{-1}] \, M_{\odot}.
\end{equation}
The value of the chirp mass that minimizes the statistic,
\begin{equation}
    \hat{\mathcal{M}}
    =
    \arg\min_{\mathcal{M}}
    {\rm NMSE}(\mathcal{M}) ,
    \label{eq:nmse_mchirp_estimator}
\end{equation}
provides a first estimate of the source chirp mass. We do
not interpret this value as a precise parameter-estimation
result, since the Viterbi track is obtained from a
semi-coherent search and may be affected by noise features,
spectral lines, or imperfect candidate isolation. However,
it provides a useful consistency check: signal-like tracks
should not only have an excess of accumulated power, but
should also be well fitted by an inspiral trajectory within
the physical mass range explored by the search.

The two statistics are therefore complementary. The
quantity $n_{\sigma}$ measures how significant the recovered
track is with respect to the noise-only Viterbi background,
while ${\rm NMSE}_{\rm min}$ (NMSE, from now on) measures how compatible the track is with the expected long-inspiral morphology. A
noise fluctuation can occasionally produce a large
accumulated power, but it is not expected to follow the
specific linear behaviour in the $(t,f^{-8/3})$ plane. On the
other hand, a low-SNR signal may not produce a very large
$n_{\sigma}$, but can still be identified through a small NMSE
if its recovered track follows the expected inspiral
evolution. For this reason, we evaluate candidates in the
two-dimensional detection-statistic plane $\left(n_{\sigma}, {\rm NMSE} \right)$, which allows us to combine statistical significance and
physical consistency in a single candidate-selection
criterion.

%--------------------------------------------------------------------------------------------------------------------------------------------------------------------------------------

\section{Application to LVK data \label{sec:Analysis}}
In this work, we present a fully operational search pipeline, end-to-end, from time-series strain to a candidate evaluation. It is designed to detect transient CWs, produced during the inspiral phase of PBHs binaries within the planetary-to-sub-solar mass range. 

Based on the Viterbi algorithm, this signal-agnostic method is capable of recovering faint, slowly evolving signals buried in data. As discussed above, the timescales associated with these long-duration inspirals differ significantly from those of both typical CBCs and standard CW signals. The lower-mass systems in this range can produce signals with negligible frequency evolution within the observing band over timescales of $\mathcal{O}(\mathrm{months})$, whereas higher-mass systems generate more rapidly evolving tracks that sweep across the band on timescales of $\mathcal{O}(\mathrm{minutes})$. Consequently, these signals are subject to different classes of noise operating on different timescales. A robust and flexible search method is therefore required to reliably distinguish signals from noise while accommodating this wide range of signal timescales across the targeted mass range.

In this section, we first describe the multi-stage structure of the pipeline, from the transformation of the strain data into the frequency domain to the identification of significant candidates. We then apply the pipeline to real data from the third observing run (O3) of LIGO Hanford, validating its performance and assessing its sensitivity through a controlled scenario with injected signals. 

\subsection{Pipeline Architecture}

\begin{figure*}[t]
    \includegraphics[width=\textwidth]{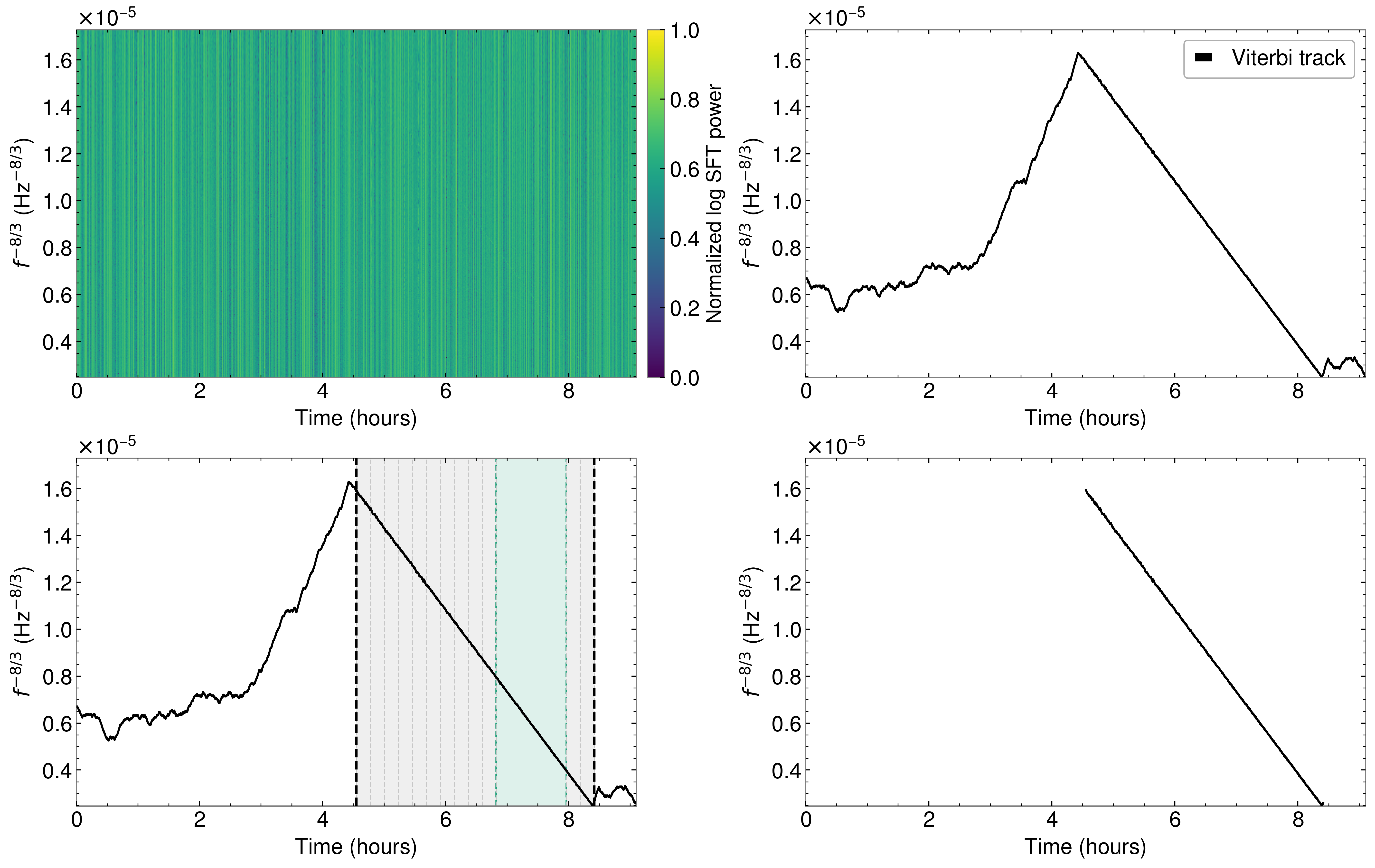}
\caption{Phases of the search pipeline applied to an injected long-inspiral signal. From top left to bottom right, the figure illustrates the different stages of the pipeline. Top left: the remapped frequency representation introduced in Sec.~\ref{sec:longinspirals}. Top right: the recovered Viterbi track, corresponding to the most likely frequency evolution path given the data.  Bottom left: the candidate-isolation procedure, which first identifies the window with the highest power significance (green) and then expands it using smaller windows on both sides. Bottom right: the final candidate, for which the pipeline outputs a pair of ranking statistics, $(n_{\sigma}, \text{NMSE})$. The injected signal, with parameters $[\mathcal{M}=10^{-2}\,M_\odot,\; d_L=80\,\mathrm{kpc}]$, is barely discernible in the spectrogram, yet its track is successfully recovered by the Viterbi algorithm.}
\label{fig:pipeline_architec}
\end{figure*}

The search method presented in this work is built around the Viterbi algorithm. However, the Viterbi algorithm constitutes only one component of a broader multi-stage analysis pipeline.

The pipeline begins by transforming the strain time series into a set of SFTs, as the analysis is performed entirely in the frequency domain. At this stage, the key parameter is the coherence time of the Fourier transform, $T_{\mathrm{SFT}}$. As discussed in Sec.~\ref{sec:sft}, Ref.~\cite{Alestas:2024ubs} showed that the optimal coherence time for signal recovery, $T_{\mathrm{SFT}}^{\rm opt}$, depends directly on the chirp mass of the system. Since the signal parameters are not known \textit{a priori} and our search targets a broad mass range, we employ multiple values of $T_{\mathrm{SFT}}$ rather than adopting a single fixed choice.

We determine the optimal $T_{\mathrm{\rm SFT}}$ values across the mass range $10^{-4} M_{\odot} \lesssim \mathcal{M} \lesssim 10^{-1} M_{\odot}$ by requiring the SNR loss, relative to the maximum achievable SNR, to remain below $1\%$~\cite{Alestas:2024ubs}. Following this criterion, we find that the optimal $T_{\mathrm{\rm SFT}}$ values span the range from $2$ to $88$ seconds, requiring thirteen distinct coherence times. As a result, the data are represented by multiple time-frequency maps rather than by a single spectrogram. This stage constitutes the primary computational bottleneck of the pipeline, although its cost can be substantially reduced, as discussed in the next section.

Once the spectrograms are built, following the variable change presented in Sec.~\ref{sec:longinspirals}, we remap the frequency representation leading to a new frequency map $(t, f^{-8/3})$, where the chirps become straight lines (see Fig.~\ref{fig:newmap}). This transformation is motivated by two main considerations. First, it allows us to exploit the transition matrix (see Sec.~\ref{sec:viterbi}) to define a prior that favors center and downward (CD) transitions, thereby enabling the Viterbi algorithm to follow the signal more effectively. Second, in this representation, the slope of the track depends only on the chirp mass of the system. This property forms the basis of the NMSE fitting statistic (see Sec.~\ref{sec:detstatistics}), which provides a powerful tool for identifying and ranking significant candidates.

Now, with a set of remapped spectrograms constructed, the Viterbi algorithm is applied independently to each of them. This step is computationally efficient and accounts for only a small fraction of the overall computational cost. For each spectrogram, the algorithm identifies the most probable track given the data. This allows us to select the optimal $T_{\mathrm{\rm SFT}}$ representation by comparing the corresponding detection statistic, $n_{\sigma}$ (see Sec.~\ref{sec:detstatistics}), as the spectrogram that best matches the signal will yield the track with the highest accumulated power.
 
After the Viterbi track has been recovered, the next step is to identify the portion of the track that is most likely associated with a physical signal. To this end, we implement a four-stage isolation procedure. First, the recovered track is divided into eight time windows. For each window, we compute the fraction of the total accumulated power ($\rho_i$) contained within it. The two windows with the largest power fractions are retained as initial signal-candidate regions and passed to the next stage. In this second phase, the track segments of these candidate windows is compared with the expected inspiral evolution in the remapped frequency coordinate, Eq.~\eqref{dotfgwnew}. A fitting is performed over the target mass range using the normalized mean-square error (NMSE), defined in Eq.~\eqref{eq:nmse_definition}. We retain the window with the lowest NMSE, corresponding to the segment that best matches the expected inspiral behaviour.

The third stage refines the selected window by allowing its boundaries to expand or contract. This is done iteratively by adding or removing small portions of the track at the left and right boundaries independently, and recomputing the NMSE after each modification. This procedure allows us to recover the full signal-containing segment of the Viterbi track, as illustrated in Fig.~\ref{fig:pipeline_architec}, and a metric evaluation of how well it fits to the expected inspiral evolution. Finally, the isolated segment is evaluated using its detection statistics $(n_{\sigma}, \text{NMSE})$. A detection threshold is defined in the ranking statistics plane, using a chosen false-alarm ratio (FAR), allowing each recovered track to be classified either as consistent with noise or as a significant signal candidate.

All in all, the pipeline combines the computational efficiency of the Viterbi algorithm with a dedicated candidate-isolation procedure tailored to long inspiral signals. It enables robust searches over the broad range of planetary-to-sub-solar PBHs, while providing a physically motivated detection statistic for candidate selection. The pipeline is available in~\cite{viterbi_pbh_github}, where all the codes can be implemented in any GW dataset.

\subsection{Results}
Now, we are ready to test the full pipeline on real data. Specifically, we apply the search method to data from the O3b LIGO Hanford~\cite{KAGRA:2013rdx, LIGOScientific:2014pky} observing run, into which we inject a population of simulated long-inspiral signals. This controlled setup allows us to evaluate the behaviour of the detection statistics over a broad set of injections and to assess the sensitivity and robustness of the method.

The dataset and injection parameters used in the analysis are summarized in Table \ref{tab:benchmark}. The chirp mass and luminosity distance are sampled logarithmically. In total, we inject approximately 600 signals into nearly 1000 hours of data. These low-mass binary inspirals are generated using the 3.5PN \texttt{TaylorT3} approximant~\cite{Alestas:2024ubs, TaylorApproximants}. This approximant is particularly convenient for the present application because it provides the possibility of generating the signal segment by segment, which is required in practice due to memory limitations.
\begin{table}[t]
\centering
\begin{tabular*}{0.95\columnwidth}{@{\extracolsep{\fill}}ll@{}}
\toprule
\multicolumn{2}{c}{\textbf{Search parameters}} \\
\toprule
Dataset & O3b \\
Detector & H1 \\
Chunk duration & $32768\ \mathrm{s}$ \\
Total observation time & $\sim 1000\ \mathrm{h}$ \\
Sampling & $512\ \mathrm{Hz}$ \\
\midrule
Chirp mass, $\mathcal{M}$ & $[10^{-4}, 10^{-1}]\,M_{\odot}$ \\
Luminosity distance, $d_L$ & $[0.1, 150]\ \mathrm{kpc}$ \\
Mass ratio, $q$ & $1$ \\
Right ascension, $\alpha$ & $[0,2\pi]$ \\
Declination, $\delta$ & $[-\pi/2,\pi/2]$ \\
Polarization, $\psi$ & $[0,\pi]$ \\
Inclination, $\iota$ & $0$ \\
\bottomrule
\end{tabular*}
\caption{Injection parameters defining the benchmark search case used in this work.}
\label{tab:benchmark}
\end{table}

For each injection, we compute the time-dependent projection of the waveform onto the Hanford detector using \texttt{LALSuite}~\cite{lalsuite}. The resulting detector-frame signals are then added to gap-free segments of O3b strain data obtained from GWOSC public dataset~\cite{O3data}.

Following the theoretical estimation of the optimal SFT length, given in Eq.~\eqref{eq:TSFTopt}, we find that the chirp mass range considered here requires a set of thirteen $T_{\rm SFT}$, ranging from 2 to 88 s. The SFTs are produced with the \texttt{MakeSFTs} routine from the \texttt{LALPulsar} library~\cite{lalsuite, swiglal}, within the optimal frequency band $[61.1,126.8]~{\rm Hz}$ (derived in Sec.~\ref{sec:sft}). 
We then use the Python \texttt{soapcw} package~\cite{soapcw} to run the Viterbi algorithm, extract the most likely tracks, and apply the full candidate-selection pipeline described in the last section.

\begin{figure}[t]
    \centering
    \includegraphics[width=\linewidth]{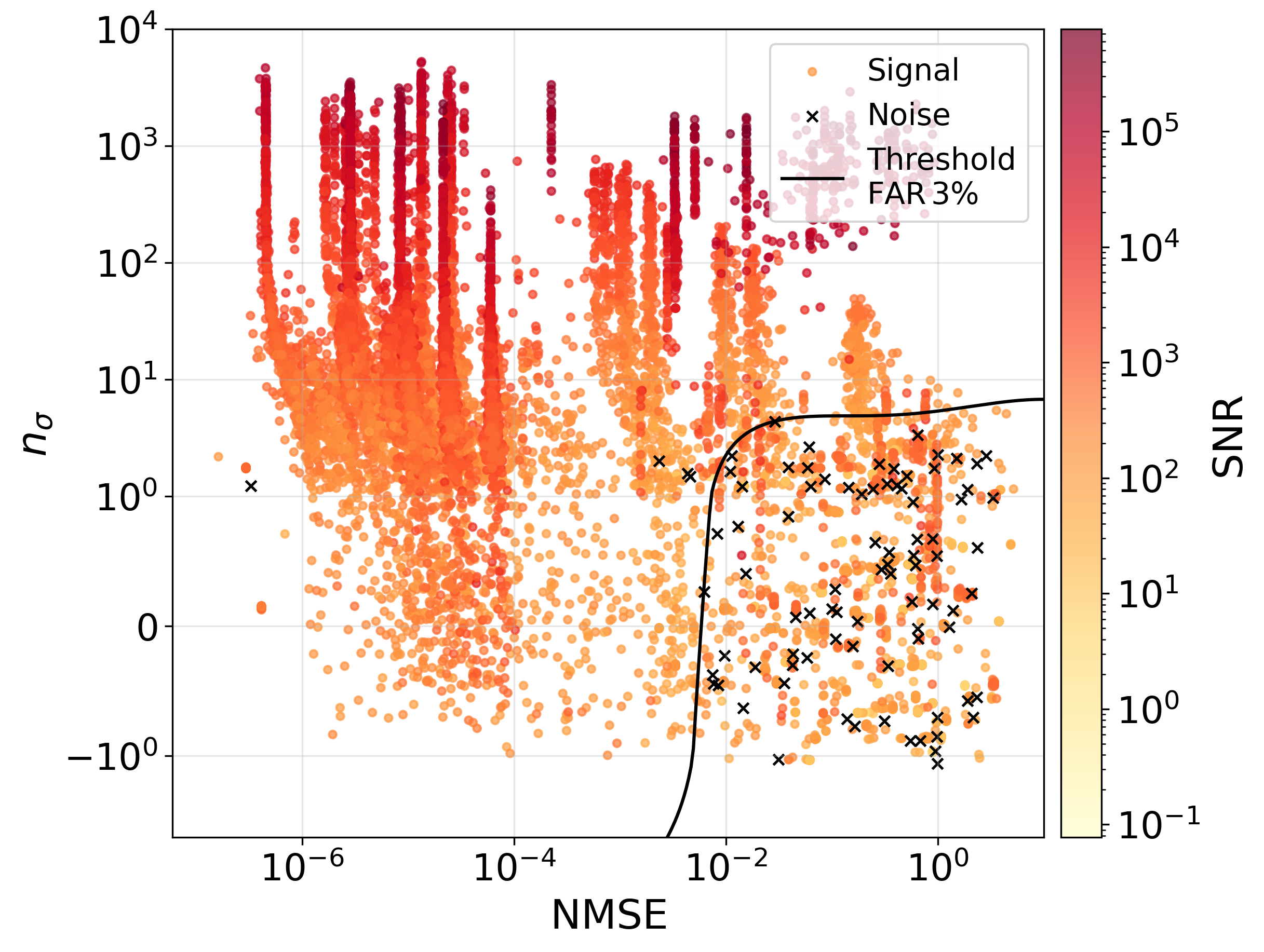}
    \caption{Triggers distribution for the search in O3b Hanford data. Each search output is represented as a point in the detection-statistic plane. Dots correspond to searches performed on data containing injected signals, with the color indicating the corresponding signal SNR, while black crosses represent searches on pure noise realizations. The noise triggers are clearly concentrated in the lower-right region of the plane, characterized by low $n_{\sigma}$ and high NMSE values, whereas signal candidates predominantly populate the complementary regions.}
    \label{fig:triggers}
\end{figure}

Figure~\ref{fig:triggers} presents one of the main results, showing the distribution of triggers recovered by the pipeline from noise-only and signal-injected data sets. We first run the search on the O3b raw data, which provide a representative sample of noise triggers (we use the word trigger as the output of a search, a point in ranking statistics plane). This allows us to characterize the background distribution in the detection statistic plane. As expected, noise triggers predominantly populate the region with large NMSE and low $n_{\sigma}$ (black cross markers in Fig.~\ref{fig:triggers}), corresponding to poor agreement with the inspiral model and low statistical significance. 
 
Then, we run the pipeline after injecting the signal population, summarized in Table \ref{tab:benchmark}, into different noise realizations. In this case, a large fraction of the recovered triggers moves towards the significant region of the plane, characterized by low NMSE and high $n_{\sigma}$.  In Fig.~\ref{fig:triggers}, the triggers recovered from signal-injected data are shown as colored dots, where the color indicates the corresponding signal SNR, computed as described in Appendix~\ref{app:SNR}.

We then define a detection threshold by fixing the false-alarm ratio to $\mathrm{FAR}=3\%$. To this end, we construct a polynomial decision boundary in the detection-statistic plane (solid black line in Fig.~\ref{fig:triggers}), optimized to maximize the number of true positives at fixed FAR. We consider polynomial functions up to cubic order, since higher-order boundaries do not provide a significant improvement in recovery performance. Future work will require a more extensive background characterization through additional searches over raw datasets to improve the significance of systematic searches.

The resulting pipeline recovers a substantial range of the injected parameter space. Nevertheless, a number of injected signals are still classified as noise. This is truly expected since there is a sensitivity limitation associated with the Viterbi reconstruction itself. Below a certain SNR, the accumulated signal power becomes comparable to, or smaller than, that of common instrumental artifacts. In this regime, the Viterbi algorithm can preferentially follow noise features or spectral lines rather than the true signal track. Additional losses arise from the candidate-isolation stage of the pipeline. Tracks that do not exhibit a well-defined concentration of power, or whose recovered morphology becomes too diffuse at low SNR, can be difficult to fit reliably with the expected inspiral evolution. These losses are expected on real data, where real noise increase the difficulty to isolate signal events.

In these low-SNR scenarios, the importance of the new metric NMSE becomes evident. While $n_{\sigma}$ is highly effective for independently identifying high-SNR candidates, its discriminating ability get worse once the accumulated signal power becomes comparable to the background fluctuations. In this regime, the NMSE plays a key role in distinguishing inspiral-like signals from noise-induced tracks.

\begin{figure}[b]
    \centering
    \includegraphics[width=\linewidth]{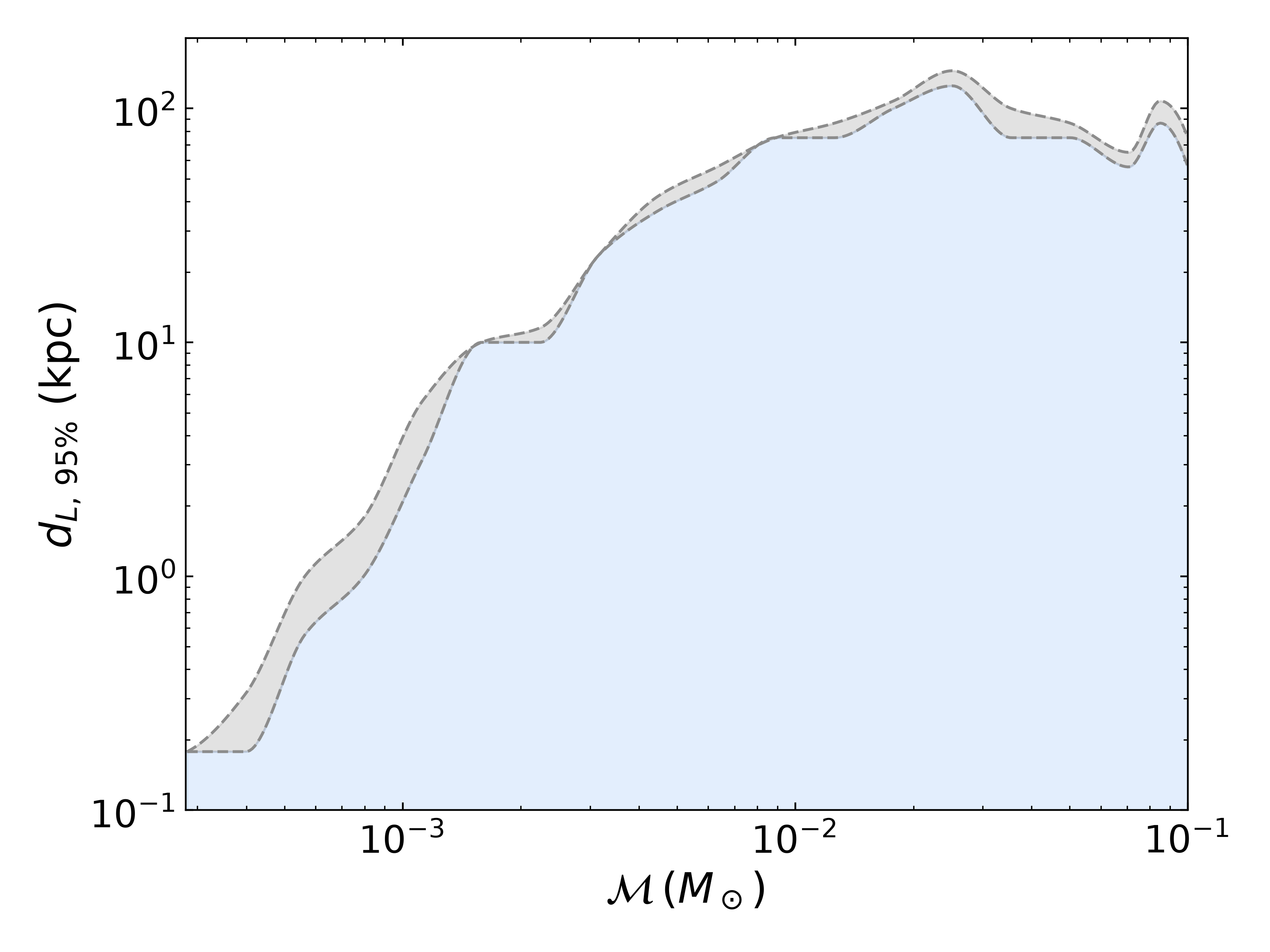}
    \caption{Distance reach curve corresponding to $d_{L,95\%}$, estimated from injection campaigns performed on 1000 hours of O3b Hanford data. For the population injected (Table \ref{tab:benchmark}), we can recover at least the $95\%$ of the signals within the blue contour. The gray shaded contour represents the $1\sigma$ confidence interval obtained using the Wilson score interval. As expected, the sensitivity increase with the chirp mass of the binary, up to a point where the signal evolves so fast that it becomes more difficult to isolate.}
    \label{fig:sensitivity}
\end{figure}

Now that we have run the search and defined the threshold, we can asses the sensitivity of our pipeline over the population mentioned in Table~\ref{tab:benchmark}. In Fig.~\ref{fig:sensitivity}, we show the distance reach as a function of chirp mass. In particular, we report the $d_{L,95\%}$ curve, defined as the luminosity distance up to which signals are recovered in at least $95\%$ of the injections.

We also show the $1\sigma$ confidence interval (gray shaded region) associated to the maximum distance reach, computed using the Wilson score interval~\cite{Wilson01061927}. This binomial estimator is particularly well suited to our case, since it remains reliable for small sample sizes and does not collapse at the extremes, unlike the standard normal approximation. It allows us to visualize the uncertainty associated with the limited number of realizations and the intrinsically random behaviour of real detector data.

For the injected population, we recover a substantial fraction of the signals across the explored parameter space. Above a chirp mass of $\mathcal{M}\gtrsim 5 \times 10^{-3} \,M_\odot$, the search reaches distances of $d_{L,95\%}\gtrsim 50$ kpc. Notably, over most of the targeted mass range, the sensitivity extends to Galactic scales. This is particularly relevant for our science goals, as the most promising candidates are expected to be compact binaries located within the Milky Way. In the most favorable region, the pipeline is able to recover signals with a $95\%$ efficiency out to distances of $135$ kpc, corresponding to chirp masses around $\mathcal{M}\sim  2 \times 10^{-2} \,M_\odot$.

An interesting trend can be observed in the high-mass range of the distance reach curve. As the chirp mass increases, the SNR also increases, leading to the expected improvement in sensitivity. However, at the higher explored mass range, the sensitivity exhibits a slightly decrease despite the continued increase in SNR. This behaviour can be understood from the morphology of the signals. In this high mass regime, the binaries evolve significantly faster, spending only minutes within the detector band. Therefore, the signals become less similar to the long transient CWs that motivate the design of our pipeline, so the candidate recovery become more challenging. The large increase in orbital frequency of these systems also limits the performance of the Viterbi reconstruction, since the transition matrix employed in this work only allows jumps of one frequency bin per step and therefore cannot fully capture such rapid frequency evolution. Nevertheless, the decrease remains moderate, and the pipeline continues to recover a large fraction of the injected population.

Finally, we evaluate the accuracy of the chirp mass estimation obtained from the NMSE minimization, Eq.~\eqref{eq:nmse_mchirp_estimator}. To quantify this rough estimate, we define the relative error as
\begin{equation}
\delta = \frac{\hat{\mathcal{M}}-\mathcal{M}_{\rm true}}
{\mathcal{M}_{\rm true}},
\end{equation}
where $\hat{\mathcal{M}}$ is the chirp mass associated with the final candidate and $\mathcal{M}_{\rm true}$ is the injected value. This normalization provides a dimensionless quantity that can be directly compared across the explored parameter space.

\begin{figure}[t]
\centering
\includegraphics[width=\columnwidth]{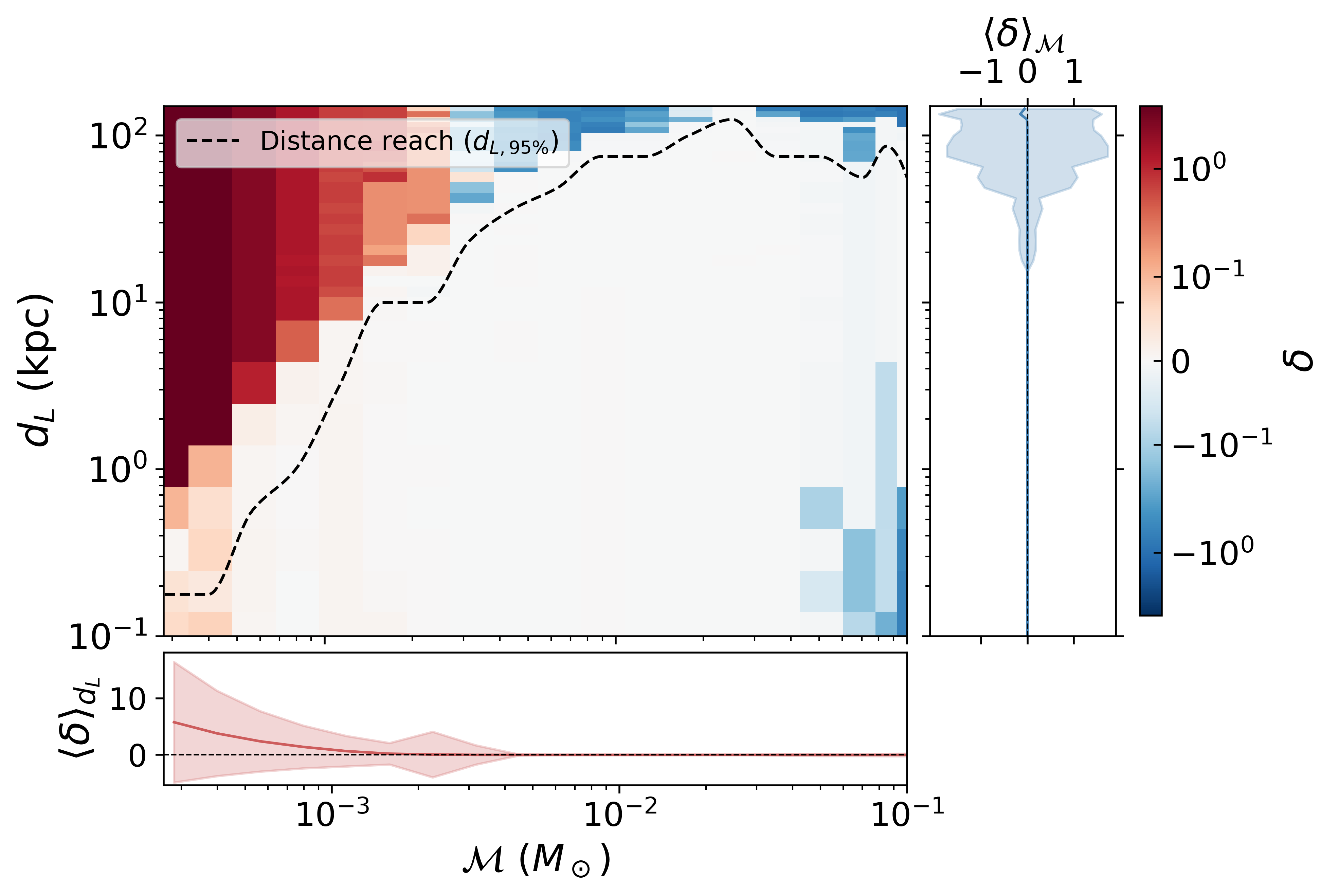}
\caption{Relative error of the chirp mass estimation obtained from the NMSE minimization. The central panel shows the error across the injected parameter space, while the bottom and right panels display the corresponding marginal distributions with an standard deviation ($1\sigma$) contour. The distance reach curve $(d_{L,95\%})$ is shown for comparison. As expected, the reconstruction remains accurate within the detectable region and progressively degrades towards lower masses and larger distances.}
\label{fig:delta}
\end{figure}

Figure~\ref{fig:delta} shows the relative error obtained for the full set of injections. Across most of the detectable region, the error remains close to zero, indicating that the pipeline is able not only to recover the signal, but also to provide a reliable estimate of its chirp mass. This behaviour closely follows the distance reach curve shown in Fig.~\ref{fig:sensitivity}: within the region where signals are efficiently recovered, the chirp-mass reconstruction remains accurate, whereas the error rapidly increases as the injections move beyond the distance reach limit.

A small deviation from this overall behaviour is observed at the highest chirp masses. As discussed previously, these signals evolve rapidly through the detector band, producing short-duration tracks that are more difficult to isolate from the background. Consequently, the fitted candidates may retain a small noise contribution, leading to modest biases in the chirp mass estimate ($|\delta| \lesssim 0.1$). Although these signals are still reliably detected, improving the candidate isolation in this regime will be important for future analyses.

We also show the marginal distributions with respect to chirp mass and luminosity distance. As expected, the relative error converges to zero throughout the high-SNR region of the parameter space and increases only for lower masses and larger distances, where the recovered tracks become progressively more affected by noise.

A small shift can be observed when comparing Fig.~\ref{fig:sensitivity} and \ref{fig:delta}. The region characterized by low relative errors extends slightly beyond the $d_{L,95\%}$ curve, even surpassing the upper limit of the Wilson confidence interval. This behaviour is expected and reflects the effect of the detection threshold: some signals close to the sensitivity boundary still produce tracks that are well fitted by the inspiral model and therefore yield accurate chirp-mass estimates, although their statistical significance is not sufficient to satisfy the selection criteria. The effect remains small, indicating that the sensitivity threshold is closely aligned with the onset of degraded parameter reconstruction.

%--------------------------------------------------------------------------------------------------------------------------------------------------------------------------------------

\section{Conclusions \label{sec:conclusions}}

%--------------------------------------------------------------------------------------------------------------------------------------------------------------------------------------
Primordial black hole binaries, in the planetary-to-subsolar mass range, have recently attracted renewed interest due to several observational evidences. Detecting such systems using gravitational waves remains challenging, as their long-lasting inspiral signals lie in the intermediate regime between conventional CBCs and CW searches.

In this work, we have developed and validated a fully operational search pipeline targeting these compact binaries during their inspiral phase. The method is based on the Viterbi algorithm, which recovers the most probable track from the data, assuming that it evolves as a Markov process. To enhance its performance, we introduce a new frequency-domain representation, $(t,f^{-8/3})$, where inspiral chirps become approximately linear, allowing us to constrain the transition matrix and reduce the signal evolution to a single parameter, the binary's chirp mass. The recovered Viterbi track is then processed through a candidate isolation procedure, where we extract a potential signal from background noise. Then, the candidate is evaluated using the detection statistics $(n_\sigma,\mathrm{NMSE})$, which quantify its significance and consistency with the expected inspiral morphology, respectively, discriminating possible signal candidates from random noise fluctuations.

Having presented the foundations of the pipeline, we validated the search method using nearly 1000 hours of O3b LIGO Hanford data. After characterizing the background distribution and defining a detection threshold corresponding to a false-alarm ratio of $\mathrm{FAR}=3\%$, we carried out a search over an injected signal population. We presented the resulting distance reach curve, specifically the $d_{L,95\%}$. It reaches Galactic scales over most of the explored mass range, extending up to luminosity distances of $135\,\mathrm{kpc}$ in the most favorable region. An interesting trend was observed in the high-mass regime, where fast-evolving binaries lead to shorter transients that are more challenging to recover. Nevertheless, the pipeline still performs relatively well, reaching distances $\gtrsim 65 \,\mathrm{kpc}$.

Furthermore, we showed that the NMSE detection statistic provides a useful first estimation of the binary's chirp mass. As it is has been shown in Fig. \ref{fig:delta}, within the sensitivity region of the search, the recovered values remain very close to the injected masses, enabling a first characterization of future signal candidates.

The search pipeline presented here is computationally efficient, with the generation of the SFTs representing the main bottleneck. Once the spectrograms are produced, a complete candidate evaluation requires only $\mathcal{O}(\mathrm{sec})$ for a 10-hour data realization. Moreover, we expect a significant increase in sensitivity when using multi-detector data, where uncorrelated noise across the network will help to differentiate between local detector glitches from possible candidates.

Future developments may include leading order corrections in the frequency evolution to account for eccentricity, a likely property of PBH binaries in this mass range. Furthermore, extending the search to longer datasets would improve the characterization of the background noise, and consequently a more robust assessment of candidate significance. It would also be interesting to perform dedicated searches over narrower chirp-mass ranges, increasing the sensitivity and reducing significantly the computational cost by limiting the number of $T_{\rm SFT}$ values considered, as SFT generation dominates the runtime and must be repeated for each coherence time. 
Finally, a systematic search of the full available observing data set, including the O4 run, will be essential to further investigate the nature of planetary-to-subsolar mass PBH binaries, whose discovery could have profound implications for our understanding of dark matter, cosmology, and fundamental physics.

\section*{Acknowledgements}
The authors would like to thank D. Keitel, P. B. Covas for useful discussions, and G. Morrás for his help in developing the code and waveform approximant as well as for his careful review of the manuscript. The authors thankfully acknowledge the computer resources at MareNostrum 5 and the technical support provided by the Barcelona Supercomputing Center (BSC) through the grant RES-AECT-2025-3-0050 from the Red Española de Supercomputación (RES). This work is partially funded by the European Commission – NextGenerationEU, through Momentum CSIC Programme: Develop Your Digital Talent. We acknowledge HPC support by Emilio Ambite, staff hired under the Generation D initiative, promoted by Red.es, an organisation attached to the Spanish Ministry for Digital Transformation and the Civil Service, for the attraction and retention of talent through grants and training contracts, financed by the Recovery, Transformation and Resilience Plan through the EU’s Next Generation funds. The authors are grateful for computational resources provided by the  LIGO Laboratory	and supported by the National Science Foundation Grants PHY-0757058 and PHY-0823459. 

The authors acknowledge support from the Grant IFT Centro de Excelencia Severo Ochoa No CEX2020-001007-S and the Strategic Network REDONGRA through AEI project RED2024-153735-E, funded by MCIN/AEI/10.13039/501100011033. R.R. is supported through the Conselleria d'Educació i Universitats del Govern de les Illes Balears via an FPI-CAIB doctoral grant ($\mathrm{FPI}\_\mathrm{2024}\_\mathrm{20}$) with funds from the European Social Fund+ in the framework of the Balearic Islands ESF+ Program 2021-2027 and by the Universitat de les Illes Balears (UIB) with funds from the Programa de Foment de la Recerca i la Innovació de la UIB 2024-2026 (supported by the yearly plan of the Tourist Stay Tax ITS2023-086); the Spanish Agencia Estatal de Investigación grants PID2022-138626NB-I00, RED2024-153978-E, RED2024-153735-E, funded by MICIU/AEI/10.13039/501100011033 and the ERDF/EU; and the Comunitat Autònoma de les Illes Balears through the Conselleria d'Educació i Universitats with funds from the European Union - European Regional Development Fund (ERDF) (SINCO2022/18146 - Plataforma HiTech-IAC3-BIO). G.A. is supported by the Spanish Research Agency’s Consolidaci\'on Investigadora 2024 grant CNS2024-154430. S.K. is supported by the I+D grant PID2023-149018NB-C42 funded by MCIN/AEI/10.13039/501100011033, the Leonardo Grant for Scientific Research and Cultural Creation 2024 from the BBVA Foundation, and Japan Society for JSPS KAKENHI Grant no. JP23H00110 and JP24K00624. J.G.B is supported by the I+D grant PID2024-159420NB-C43 funded by MCIN/AEI/10.13039/501100011033.

This material is based upon work supported by NSF's LIGO Laboratory which is a major facility fully funded by the National Science Foundation. This research has made use of data or software obtained from the Gravitational Wave Open Science Center (gwosc.org), a service of the LIGO Scientific Collaboration, the Virgo Collaboration, and KAGRA. This material is based upon work supported by NSF's LIGO Laboratory which is a major facility fully funded by the National Science Foundation, as well as the Science and Technology Facilities Council (STFC) of the United Kingdom, the Max-Planck-Society (MPS), and the State of Niedersachsen/Germany for support of the construction of Advanced LIGO and construction and operation of the GEO600 detector. Additional support for Advanced LIGO was provided by the Australian Research Council. Virgo is funded, through the European Gravitational Observatory (EGO), by the French Centre National de Recherche Scientifique (CNRS), the Italian Istituto Nazionale di Fisica Nucleare (INFN) and the Dutch Nikhef, with contributions by institutions from Belgium, Germany, Greece, Hungary, Ireland, Japan, Monaco, Poland, Portugal, Spain. KAGRA is supported by Ministry of Education, Culture, Sports, Science and Technology (MEXT), Japan Society for the Promotion of Science (JSPS) in Japan; National Research Foundation (NRF) and Ministry of Science and ICT (MSIT) in Korea; Academia Sinica (AS) and National Science and Technology Council (NSTC) in Taiwan.

\appendix
\section{SNR for long inspirals} 
\label{app:SNR}
\begin{figure}[t]
\centering
\includegraphics[width=\columnwidth]{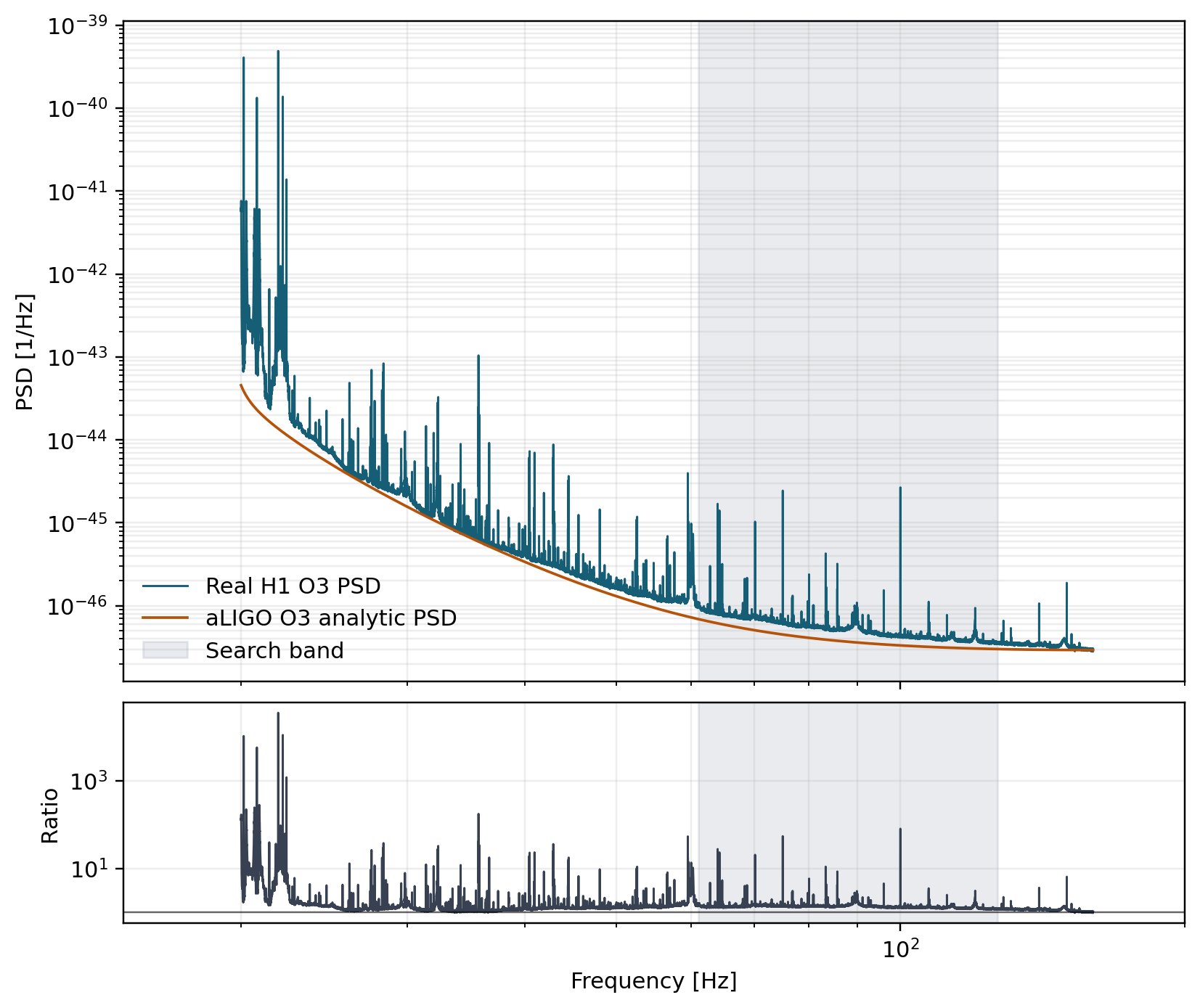}
\caption{Average noise power spectral density obtained from O3b Hanford data. We compute the PSD with $1000$ hours of O3b data using Welch's method (blue line), compared with the analytic detector sensitivity curve \texttt{aLIGOO3LowT1800545} (orange line). The real data PSD preserves spectral structures and instrumental lines present in the interferometer data.
}
\label{fig}
\end{figure}

In this section, we describe how the signal-to-noise ratio (SNR) of the injected long-inspiral signals is computed. As shown in Fig.~\ref{fig}, the injections span a broad range of SNR values, depending on the source parameters, detector response and noise realization. Since the analysis is performed on real interferometric data, a realistic estimate of the detector noise power spectral density (PSD) is required.

For this purpose, we construct an average PSD directly from O3b LIGO Hanford data. The strain data correspond to the public GWOSC dataset \texttt{H1:GWOSC-4KHZ\_R1\_STRAIN} from the O3b observing run~\cite{KAGRA:2023pio}. Rather than using an analytical sensitivity curve, we estimate the noise spectrum from the same dataset used throughout the analysis. This allows the SNR calculation to account for the actual noise properties of the detector, including spectral lines and deviations from ideal stationary Gaussian noise.

For each O3b strain realization, we estimate the PSD using Welch's method. The strain time series is divided into segments of duration $T_{\rm seg}=512\,{\rm s}$ with a 50$\%$ overlap, corresponding to a stride of $256\,{\rm s}$. A Hann window is applied to each segment before computing its discrete Fourier transform. The one-sided PSD of each segment is then obtained from the corresponding periodogram. To reduce the impact of glitches and narrow spectral artifacts, the individual periodograms are combined using the median estimator rather than the arithmetic mean.

The resulting PSD estimate for each realization, $\hat S^{(m)}(f)$, is then averaged over all segments,
\begin{equation}
\bar S(f)=\frac{1}{M}\sum_{m=1}^{M}\hat S^{(m)}(f),
\end{equation}
where $M$ denotes the total number of independent noise realizations. This ensemble average provides a robust estimate of the detector noise spectrum used throughout the analysis.

Figure~\ref{fig} shows the resulting average PSD together with the analytical PSD (\texttt{aLIGOO3LowT1800545}), the expected sensitivity for the Hanford detector during O3. The agreement is generally good across the frequency range of interest, while the real-data PSD naturally retains the spectral lines present in the detector data and an expected shift respect to the ideal behaviour. For this reason, we use $\bar S(f)$ in all SNR calculations presented in this work.

Once the detector-frame waveform has been generated, the signal is projected onto the Hanford interferometer according to
\begin{equation}
h(t)=F_+ h_+(t)+F_\times h_\times(t),
\end{equation}
where $F_+$ and $F_\times$ are the detector antenna pattern functions~\cite{10.1093/mnras/224.1.131}. The corresponding Fourier-domain waveform $\tilde h(f)$ is then used to compute the optimal matched-filter SNR.

For a single data frame, the optimal SNR is defined as
\begin{equation}
\rho^2_{\rm frame} = \langle h | h \rangle =  4\int_{f_{\rm low}}^{f_{\rm high}}
\frac{|\tilde h(f)|^2}
{\bar S(f)}
,df,
\end{equation}
where the integration is restricted to the optimal frequency interval derived in Sec.~\ref{sec:sft}, $[61.1,126.8]~{\rm Hz}$.

In practice, this integral is evaluated numerically on the discrete frequency grid of the waveform, after linearly interpolating the PSD to the corresponding frequency bins. Since the inspiral signal extends over several consecutive data frames, the total SNR is obtained by incoherently summing the SNR squared from all analyzed frames,
\begin{equation}
\rho^2_{\rm total} =\sum_{k=1}^{N_{\rm frames}}
\rho_k^{2},
\end{equation}
where the number of analyzed frames is set to $N_{\rm frames}=8$ in the benchmark searches presented here. 

The final optimal SNR reported throughout the paper is therefore
\begin{equation}
\rho_{\rm opt} = \sqrt{\rho^2_{\rm total} }
\end{equation}
This quantity is used in Fig.~\ref{fig} to characterize the injected population and to relate the recovered detection statistics to the intrinsic strength of the signal.

\bibliographystyle{apsrev4-2}
\bibliography{Bibliography}

@ARTICLE{Viterbi,
  author={Viterbi, A.},
  journal={IEEE Transactions on Information Theory}, 
  title={Error bounds for convolutional codes and an asymptotically optimum decoding algorithm}, 
  year={1967},
  volume={13},
  number={2},
  pages={260-269},
  doi={10.1109/TIT.1967.1054010}}

@misc{lalsuite,
       author         = "{LIGO Scientific Collaboration} and {Virgo Collaboration} and {KAGRA Collaboration}",
       title          = "{LVK} {A}lgorithm {L}ibrary - {LALS}uite",
       howpublished   = "Free software (GPL)",
       doi            = "10.7935/GT1W-FZ16",
       year           = "2018"
}

@article{swiglal,
          title     = "{SWIGLAL: Python and Octave interfaces to the LALSuite gravitational-wave data analysis libraries}",
          author    = "Karl Wette",
          journal   = "SoftwareX",
          volume    = "12",
          pages     = "100634",
          year      = "2020",
          doi       = "10.1016/j.softx.2020.100634"
}

@article{Clesse:2016vqa,
    author = "Clesse, Sebastien and Garc\'\i{}a-Bellido, Juan",
    title = "{The clustering of massive Primordial Black Holes as Dark Matter: measuring their mass distribution with Advanced LIGO}",
reportNumber = "TTK-16-10, IFT-UAM-CSIC-16-027",
    doi = "10.1016/j.dark.2016.10.002",
    journal = "Phys. Dark Univ.",
    volume = "15",
    pages = "142--147",
    year = "2017"
}

@article{Magee:2018opb,
    author = "Magee, Ryan and Deutsch, Anne-Sylvie and McClincy, Phoebe and Hanna, Chad and Horst, Christian and Meacher, Duncan and Messick, Cody and Shandera, Sarah and Wade, Madeline",
    title = "{Methods for the detection of gravitational waves from subsolar mass ultracompact binaries}",
reportNumber = "LIGO-DCC-P1800231-v3",
    doi = "10.1103/PhysRevD.98.103024",
    journal = "Phys. Rev. D",
    volume = "98",
    number = "10",
    pages = "103024",
    year = "2018"
}

@article{Carr_2026,
   title={Primordial black holes: constraints, potential evidence and prospects},
   volume={49},
   ISSN={1826-9850},
   url={http://dx.doi.org/10.1007/s40766-026-00080-z},
   DOI={10.1007/s40766-026-00080-z},
   number={5},
   journal={La Rivista del Nuovo Cimento},
   publisher={Springer Science and Business Media LLC},
   author={Carr, Bernard and Iovino, Antonio J. and Perna, Gabriele and Vaskonen, Ville and Veermäe, Hardi},
   year={2026},
   month=Mar, pages={225–274} }

@article{Abbott_2023,
    author = "Abbott, R. and others",
    collaboration = "KAGRA, VIRGO, LIGO Scientific",
    title = "{Population of Merging Compact Binaries Inferred Using Gravitational Waves through GWTC-3}",
reportNumber = "LIGO-P2100239 ; Data release: https://zenodo.org/record/5655785, LIGO-P2100239",
    doi = "10.1103/PhysRevX.13.011048",
    journal = "Phys. Rev. X",
    volume = "13",
    number = "1",
    pages = "011048",
    year = "2023"
}

@article{Andres-Carcasona:2023zny,
    author = "Andr{\'e}s-Carcasona, Marc and Piccinni, Ornella Juliana and Mart{\'\i}nez, Mario and Mir, Lluisa-Maria",
    title = "{BSD-COBI: New search pipeline to target inspiraling light dark compact objects.}",
    doi = "10.22323/1.449.0067",
    journal = "PoS",
    volume = "EPS-HEP2023",
    pages = "067",
    year = "2024"
}

@article{Wyrzykowski_2011,
   title={The OGLE view of microlensing towards the Magellanic Clouds - IV. OGLE-III SMC data and final conclusions on MACHOs★: The OGLE-III view of microlensing towards the SMC},
   volume={416},
   ISSN={0035-8711},
   url={http://dx.doi.org/10.1111/j.1365-2966.2011.19243.x},
   DOI={10.1111/j.1365-2966.2011.19243.x},
   number={4},
   journal={Monthly Notices of the Royal Astronomical Society},
   publisher={Oxford University Press (OUP)},
   author={Wyrzykowski, L. and Skowron, J. and Kozłowski, S. and Udalski, A. and Szymański, M. K. and Kubiak, M. and Pietrzyński, G. and Soszyński, I. and Szewczyk, O. and Ulaczyk, K. and Poleski, R. and Tisserand, P.},
   year={2011},
   month=Aug, pages={2949–2961} }

@unpublished{sugiyama2026microlensingconstraintsprimordialblack,
      title={Microlensing constraints on Primordial Black Hole abundance with Subaru Hyper Suprime-Cam observations of Andromeda}, 
      author={Sunao Sugiyama and Masahiro Takada and Naoki Yasuda and Nozomu Tominaga},
      year={2026},
      eprint={2602.05840},
      archivePrefix={arXiv},
      primaryClass={astro-ph.CO},
      url={https://arxiv.org/abs/2602.05840},
      note={arXiv e-print},
}

@article{Dom_nech_2022,
   title={NANOGrav hints on planet-mass primordial black holes},
   volume={65},
   ISSN={1869-1927},
   url={http://dx.doi.org/10.1007/s11433-021-1839-6},
   DOI={10.1007/s11433-021-1839-6},
   number={3},
   journal={Science China Physics, Mechanics \& Astronomy},
   publisher={Springer Science and Business Media LLC},
   author={Domènech, Guillem and Pi, Shi},
   year={2022},
   month=Jan }

@article{Hawkins:2025mlo,
    author = "Hawkins, M. R. S. and Garc{\'\i}a-Bellido, J.",
    title = "{A critical analysis of the recent OGLE limits on stellar mass primordial black holes in the halo of the Milky Way}",
    eprint = "2509.05400",
    archivePrefix = "arXiv",
    primaryClass = "astro-ph.GA",
    reportNumber = "IFT-UAM/CSIC-25-91",
    doi = "10.1093/mnras/staf1826",
    journal = "Mon. Not. Roy. Astron. Soc.",
    volume = "544",
    number = "2",
    pages = "1950--1957",
    year = "2025"
}

@article{Garcia-Bellido:2017xfd,
  author = {García-Bellido, Juan and Clesse, Sébastien},
  title = {Constraints from microlensing experiments on clustered primordial black holes},
  journal = {Phys. Dark Univ.},
  volume = {19},
  pages = {144-148},
  year = {2018},
  doi = {10.1016/j.dark.2017.10.001}
}

@article{Gorton:2022fyb,
    author = "Gorton, Matthew and Green, Anne M.",
    title = "{Effect of clustering on primordial black hole microlensing constraints}",
    eprint = "2203.04209",
    archivePrefix = "arXiv",
    primaryClass = "astro-ph.CO",
    doi = "10.1088/1475-7516/2022/08/035",
    journal = "JCAP",
    volume = "08",
    number = "08",
    pages = "035",
    year = "2022"
}

@article{Agazie_2023,
   title={The NANOGrav 15 yr Data Set: Evidence for a Gravitational-wave Background},
   volume={951},
   ISSN={2041-8213},
   url={http://dx.doi.org/10.3847/2041-8213/acdac6},
   DOI={10.3847/2041-8213/acdac6},
   number={1},
   journal={The Astrophysical Journal Letters},
   publisher={American Astronomical Society},
   author={NANOGrav Collaboration},
   year={2023},
   month=jun, pages={L8} }

@article{Tisserand_2007,
   title={Limits on the Macho content of the Galactic Halo from the EROS-2 Survey of the Magellanic Clouds},
   volume={469},
   ISSN={1432-0746},
   url={http://dx.doi.org/10.1051/0004-6361:20066017},
   DOI={10.1051/0004-6361:20066017},
   number={2},
   journal={Astronomy \& Astrophysics},
   publisher={EDP Sciences},
   author={Tisserand, P. and others},
   year={2007},
   month=Apr, pages={387–404} }

@unpublished{domenech2026unifiedoriginprimordialblack,
      title={A Unified Origin of Primordial Black Hole Dark Matter and Nanohertz Gravitational Waves}, 
      author={Guillem Domènech and Shi Pi and Ao Wang},
      year={2026},
      eprint={2602.24061},
      archivePrefix={arXiv},
      primaryClass={astro-ph.CO},
      url={https://arxiv.org/abs/2602.24061},
      note={arXiv e-print},
}

@article{Belotsky_2019,
   title={Clusters of Primordial Black Holes},
   volume="79",
   ISSN={1434-6052},
   DOI={10.1140/epjc/s10052-019-6741-4},
   number={3},
   journal={The European Physical Journal C},
   publisher={Springer Science and Business Media LLC},
   author={Belotsky, Konstantin M. and others},
   year={2019},
   month=Mar }

@article{Morras_2023,
    author = "Morr\'as, Gonzalo and others",
    title = "{Analysis of a subsolar-mass compact binary candidate from the second observing run of Advanced LIGO}",
doi = "10.1016/j.dark.2023.101285",
    journal = "Phys. Dark Univ.",
    volume = "42",
    pages = "101285",
    year = "2023"
}

@article{Prunier_2023,
    author = "Prunier, Marine and Morr{\'a}s, Gonzalo and Siles, Jos{\'e} Francisco Nu{\~n}o and Clesse, Sebastien and Garc{\'\i}a-Bellido, Juan and Ruiz Morales, Ester",
    title = "{Analysis of the subsolar-mass black hole candidate SSM200308 from the second part of the third observing run of Advanced LIGO-Virgo}",
doi = "10.1016/j.dark.2024.101582",
    journal = "Phys. Dark Univ.",
    volume = "46",
    pages = "101582",
    year = "2024"
}

@article{Miller_2020,
   title={The Low Effective Spin of Binary Black Holes and Implications for Individual Gravitational-wave Events},
   volume={895},
   ISSN={1538-4357},
   url={http://dx.doi.org/10.3847/1538-4357/ab80c0},
   DOI={10.3847/1538-4357/ab80c0},
   number={2},
   journal={The Astrophysical Journal},
   publisher={American Astronomical Society},
   author={Miller, Simona and Callister, Thomas A. and Farr, Will M.},
   year={2020},
   month=jun, pages={128} }

@unpublished{LIGOScientific:2026wfs,
    author = "Abac, A. G. and others",
    collaboration = "LIGO Scientific, VIRGO, KAGRA",
    title = "{GWTC-5.0: Observations from the Second Part of the Fourth LIGO-Virgo-KAGRA Observing Run and Updates to the Gravitational-Wave Transient Catalog}",
    eprint = "2605.27225",
    archivePrefix = "arXiv",
    primaryClass = "gr-qc",
    reportNumber = "LIGO-P2600152",
    month = may,
    year = "2026",
    note = {arXiv e-print}
}

@unpublished{green2024primordialblackholesdark,
      title={Primordial Black Holes as a dark matter candidate -- a brief overview}, 
      author={Anne M. Green},
      year={2024},
      eprint={2402.15211},
      archivePrefix={arXiv},
      primaryClass={astro-ph.CO},
      url={https://arxiv.org/abs/2402.15211},
      note={arXiv e-print},
}

@article{Carr:2023tpt,
    author = "Carr, Bernard and Clesse, Sebastien and Garc\'ia-Bellido, Juan and Hawkins, Michael and Kuhnel, Florian",
    title = "{Observational evidence for primordial black holes: A positivist perspective}",
doi = "10.1016/j.physrep.2023.11.005",
    journal = "Phys. Rept.",
    volume = "1054",
    pages = "1--68",
    year = "2024"
}

@article{Niikura:2019kqi,
    author = "Niikura, Hiroko and Takada, Masahiro and Yokoyama, Shuichiro and Sumi, Takahiro and Masaki, Shogo",
    title = "{Constraints on Earth-mass primordial black holes from OGLE 5-year microlensing events}",
doi = "10.1103/PhysRevD.99.083503",
    journal = "Phys. Rev. D",
    volume = "99",
    number = "8",
    pages = "083503",
    year = "2019"
}

@misc{soapcw,
  author = {Bayley, Joe},
  title = {{soapcw}: a Viterbi-based search for continuous gravitational waves},
  howpublished = {\url{https://pypi.org/project/soapcw/}},
  year = {2023}
}

@article{LIGOScientific:2018mvr,
    author = "Abbott, B. P. and others",
    collaboration = "LIGO Scientific, Virgo",
    title = "{GWTC-1: A Gravitational-Wave Transient Catalog of Compact Binary Mergers Observed by LIGO and Virgo during the First and Second Observing Runs}",
reportNumber = "LIGO-P1800307",
    doi = "10.1103/PhysRevX.9.031040",
    journal = "Phys. Rev. X",
    volume = "9",
    number = "3",
    pages = "031040",
    year = "2019"
}

@article{LIGOScientific:2020ibl,
    author = "Abbott, R. and others",
    collaboration = "LIGO Scientific, Virgo",
    title = "{GWTC-2: Compact Binary Coalescences Observed by LIGO and Virgo During the First Half of the Third Observing Run}",
reportNumber = "P2000061",
    doi = "10.1103/PhysRevX.11.021053",
    journal = "Phys. Rev. X",
    volume = "11",
    pages = "021053",
    year = "2021"
}

@unpublished{LIGOScientific:2021usb,
	author = "Abbott, R. and others",
	collaboration = "LIGO Scientific, VIRGO",
	title = "{GWTC-2.1: Deep Extended Catalog of Compact Binary Coalescences Observed by LIGO and Virgo During the First Half of the Third Observing Run}",
	eprint = "2108.01045",
	archivePrefix = "arXiv",
	primaryClass = "gr-qc",
	reportNumber = "LIGO-P2100063",
	month = aug,
	year = "2021",
    note = {arXiv e-print}
}

@article{Garcia-Bellido:1996mdl,
    author = "Garc\'ia-Bellido, Juan and Linde, Andrei D. and Wands, David",
    title = "{Density perturbations and black hole formation in hybrid inflation}",
reportNumber = "SU-ITP-96-20, SUSSEX-AST-96-5-1",
    doi = "10.1103/PhysRevD.54.6040",
    journal = "Phys. Rev. D",
    volume = "54",
    pages = "6040--6058",
    year = "1996"
}

@article{Garcia-Bellido:2019vlf,
    author = "Garc{\'\i}a-Bellido, Juan and Carr, Bernard and Clesse, Sebastien",
    title = "{Primordial Black Holes and a Common Origin of Baryons and Dark Matter}",
doi = "10.3390/universe8010012",
    journal = "Universe",
    volume = "8",
    number = "1",
    pages = "12",
    year = "2021"
}

@article{Carr:2020gox,
    author = "Carr, Bernard and Kohri, Kazunori and Sendouda, Yuuiti and Yokoyama, Jun'ichi",
    title = "{Constraints on primordial black holes}",
reportNumber = "RESCEU-03/20; KEK-Cosmo-249; KEK-TH-2199; IPMU20-0024",
    doi = "10.1088/1361-6633/ac1e31",
    journal = "Rept. Prog. Phys.",
    volume = "84",
    number = "11",
    pages = "116902",
    year = "2021"
}

@article{Carr:2016drx,
    author = "Carr, Bernard and Kuhnel, Florian and Sandstad, Marit",
    title = "{Primordial Black Holes as Dark Matter}",
reportNumber = "NORDITA-2016-83",
    doi = "10.1103/PhysRevD.94.083504",
    journal = "Phys. Rev. D",
    volume = "94",
    number = "8",
    pages = "083504",
    year = "2016"
}

@unpublished{LIGOScientific:2026plm, author = "Abac, A. G. and others", collaboration = "LIGO Scientific, VIRGO, KAGRA", title = "{All-sky Searches for Continuous Gravitational Waves from Isolated Neutron Stars in the Data from the First Part of the Fourth LIGO-Virgo-KAGRA Observing Run}", eprint = "2603.14168", archivePrefix = "arXiv", primaryClass = "gr-qc", reportNumber = "LIGO-P2500416", month = mar, year = "2026",
    note = {arXiv e-print}
}

@article{SajithMenon:2025vpl, author = "Sajith Menon, Sandhya and others", title = "{GFH-v2 pipeline for searches of long-transient gravitational waves from newborn magnetars}",doi = "10.1103/jqf5-hbzk", journal = "Phys. Rev. D", volume = "113", number = "12", pages = "123042", year = "2026" }

@unpublished{Miller:2025ote, author = "Miller, Andrew L. and Pierini, Lorenzo", title = "{BinaryGFH-v2: Improved method to search for gravitational waves from sub-solar-mass, ultra-compact binaries using the Generalized Frequency-Hough Transform}", eprint = "2512.10539", archivePrefix = "arXiv", primaryClass = "gr-qc", month = dec, year = "2025",
    note = {arXiv e-print}
}

@unpublished{Wang:2025fmh, author = "Wang, Zi-Xuan and Cheng, Gong and Chen, Ju and Guo, Huai-Ke and Miller, Andrew L.", title = "{Methods for Detecting Gravitational Waves from mini-Extreme-Mass-Ratio Inspirals I: Statistics Based on Time-Frequency Signal Tracks}", eprint = "2512.21738", archivePrefix = "arXiv", primaryClass = "gr-qc", month = dec, year = "2025",
    note = {arXiv e-print}
}

@unpublished{Wang:2025lpj, author = "Wang, Zi-Xuan and Chen, Xing-Yu and Chen, Ju and Cheng, Gong and Guo, Huai-Ke and Miller, Andrew L.", title = "{Methods for Detecting Gravitational Waves from mini-Extreme-Mass-Ratio Inspirals II: A Spectral-Leakage-Aware Framework}", eprint = "2512.21739", archivePrefix = "arXiv", primaryClass = "gr-qc", month = dec, year = "2025", note = {arXiv e-print} }

@article{Miller:2024fpo, author = "Miller, Andrew L. and Aggarwal, Nancy and Clesse, S{\'e}bastien and De Lillo, Federico and Sachdev, Surabhi and Astone, Pia and Palomba, Cristiano and Piccinni, Ornella J. and Pierini, Lorenzo", title = "{Gravitational Wave Constraints on Planetary-Mass Primordial Black Holes Using LIGO O3a Data}",doi = "10.1103/PhysRevLett.133.111401", journal = "Phys. Rev. Lett.", volume = "133", number = "11", pages = "111401", year = "2024" }

@article{PhysRevD.109.022001,
  title = {GWTC-2.1: Deep extended catalog of compact binary coalescences observed by LIGO and Virgo during the first half of the third observing run},
  author = {{The LIGO Scientific Collaboration and the Virgo Collaboration}},
  journal = {Phys. Rev. D},
  volume = {109},
  issue = {2},
  pages = {022001},
  numpages = {45},
  year = {2024},
  month = {Jan},
  publisher = {American Physical Society},
  doi = {10.1103/PhysRevD.109.022001},
  url = {https://link.aps.org/doi/10.1103/PhysRevD.109.022001}
}

@article{PhysRevX.13.041039,
  title = {GWTC-3: Compact Binary Coalescences Observed by LIGO and Virgo during the Second Part of the Third Observing Run},
  author = {{LIGO Scientific Collaboration, Virgo Collaboration, and KAGRA Collaboration}},
  journal = {Phys. Rev. X},
  volume = {13},
  issue = {4},
  pages = {041039},
  numpages = {89},
  year = {2023},
  month = {Dec},
  publisher = {American Physical Society},
  doi = {10.1103/PhysRevX.13.041039},
  url = {https://link.aps.org/doi/10.1103/PhysRevX.13.041039}
}

@article{s551-7pch,
  title = {Search for planetary-mass ultracompact binaries using data from the first part of the LIGO--Virgo--KAGRA fourth observing run},
  author = {{Abac, A. G. and others}},
  journal = {Phys. Rev. D},
  pages = {},
  year = {2026},
  month = {May},
  publisher = {American Physical Society},
  doi = {10.1103/s551-7pch},
  url = {https://link.aps.org/doi/10.1103/s551-7pch}
}

@article{Abac_2026,
doi = {10.3847/2041-8213/ae2c74},
url = {https://doi.org/10.3847/2041-8213/ae2c74},
year = {2026},
month = {jun},
publisher = {The American Astronomical Society},
volume = {1004},
number = {2},
pages = {L22},
author = {{The LIGO Scientific Collaboration, the Virgo Collaboration, and the KAGRA Collaboration}},
title = {GWTC-4.0: Updating the Gravitational-wave Transient Catalog with Observations from the First Part of the Fourth LIGO–Virgo–KAGRA Observing Run},
journal = {The Astrophysical Journal Letters}
}

@article{Wilson01061927,
author = {Edwin B. Wilson},
title = {Probable Inference, the Law of Succession, and Statistical Inference},
journal = {Journal of the American Statistical Association},
volume = {22},
number = {158},
pages = {209--212},
year = {1927},
publisher = {Taylor \& Francis},
doi = {10.1080/01621459.1927.10502953},
URL = { 
        https://www.tandfonline.com/doi/abs/10.1080/01621459.1927.10502953
}
}

@article{O3data,
doi = {10.3847/1538-4365/acdc9f},
url = {https://doi.org/10.3847/1538-4365/acdc9f},
year = {2023},
month = {jul},
publisher = {The American Astronomical Society},
volume = {267},
number = {2},
pages = {29},
author = {{The LIGO Scientific Collaboration, the Virgo Collaboration, and the KAGRA Collaboration}},
title = {Open Data from the Third Observing Run of LIGO, Virgo, KAGRA, and GEO},
journal = {The Astrophysical Journal Supplement Series}
}

@article{Miller:2024jpo, author = "Miller, Andrew L. and Aggarwal, Nancy and Clesse, Sebastien and De Lillo, Federico and Sachdev, Surabhi and Astone, Pia and Palomba, Cristiano and Piccinni, Ornella J. and Pierini, Lorenzo", title = "{Method to search for inspiraling planetary-mass ultracompact binaries using the generalized frequency-Hough transform in LIGO O3a data}",doi = "10.1103/PhysRevD.110.082004", journal = "Phys. Rev. D", volume = "110", number = "8", pages = "082004", year = "2024" }

@article{LIGOScientific:2014pky,
    author = "Aasi, J. and others",
    collaboration = "LIGO Scientific",
    title = "{Advanced LIGO}",
doi = "10.1088/0264-9381/32/7/074001",
    journal = "Class. Quant. Grav.",
    volume = "32",
    pages = "074001",
    year = "2015"
}

@article{KAGRA:2013rdx,
    author = "Abbott, B. P. and others",
    collaboration = "KAGRA, LIGO Scientific, Virgo",
    title = "{Prospects for observing and localizing gravitational-wave transients with Advanced LIGO, Advanced Virgo and KAGRA}",
reportNumber = "LIGO-P1200087, VIR-0288A-12, JGW-P1808427",
    doi = "10.1007/s41114-020-00026-9",
    journal = "Living Rev. Rel.",
    volume = "19",
    pages = "1",
    year = "2016"
}

@unpublished{theligoscientificcollaboration2026searchesbinarymergerssubsolar,
      title={Searches for Binary Mergers with Sub-solar Mass Components in Data from the First Part of LIGO--Virgo--KAGRA's Fourth Observing Run}, 
      author={{The LIGO Scientific Collaboration and the Virgo Collaboration and the KAGRA Collaboration}},
      year={2026},
      eprint={2605.05444},
      archivePrefix={arXiv},
      primaryClass={astro-ph.HE},
      url={https://arxiv.org/abs/2605.05444},
      note={arXiv e-print},
}

@article{Carr:1974nx,
    author = "Carr, Bernard J. and Hawking, S. W.",
    title = "{Black holes in the early Universe}",
    doi = "10.1093/mnras/168.2.399",
    journal = "Mon. Not. Roy. Astron. Soc.",
    volume = "168",
    pages = "399--415",
    year = "1974"
}

@article{Hawking:1971ei,
    author = "Hawking, Stephen",
    title = "{Gravitationally collapsed objects of very low mass}",
    doi = "10.1093/mnras/152.1.75",
    journal = "Mon. Not. Roy. Astron. Soc.",
    volume = "152",
    pages = "75",
    year = "1971"
}

@article{KAGRA:2022dwb,
    author = "Abbott, R. and others",
    collaboration = "KAGRA, LIGO Scientific, VIRGO",
    title = "{All-sky search for continuous gravitational waves from isolated neutron stars using Advanced LIGO and Advanced Virgo O3 data}",
reportNumber = "LIGO-P2100367",
    doi = "10.1103/PhysRevD.106.102008",
    journal = "Phys. Rev. D",
    volume = "106",
    number = "10",
    pages = "102008",
    year = "2022"
}

@article{Suvorova:2016rdc,
    author = "Suvorova, S. and Sun, L. and Melatos, A. and Moran, W. and Evans, R. J.",
    title = "{Hidden Markov model tracking of continuous gravitational waves from a neutron star with wandering spin}",
doi = "10.1103/PhysRevD.93.123009",
    journal = "Phys. Rev. D",
    volume = "93",
    number = "12",
    pages = "123009",
    year = "2016"
}

@article{LIGOScientific:2016aoc,
    author = "Abbott, B. P. and others",
    collaboration = "LIGO Scientific, Virgo",
    title = "{Observation of Gravitational Waves from a Binary Black Hole Merger}",
reportNumber = "LIGO-P150914",
    doi = "10.1103/PhysRevLett.116.061102",
    journal = "Phys. Rev. Lett.",
    volume = "116",
    number = "6",
    pages = "061102",
    year = "2016"
}

@article{Bayley:2019bcb,
    author = "Bayley, Joe and Woan, Graham and Messenger, Chris",
    title = "{Generalized application of the Viterbi algorithm to searches for continuous gravitational-wave signals}",
doi = "10.1103/PhysRevD.100.023006",
    journal = "Phys. Rev. D",
    volume = "100",
    number = "2",
    pages = "023006",
    year = "2019"
}

@article{Bayley:2020zfa,
    author = "Bayley, Joseph and Messenger, Chris and Woan, Graham",
    title = "{Robust machine learning algorithm to search for continuous gravitational waves}",
doi = "10.1103/PhysRevD.102.083024",
    journal = "Phys. Rev. D",
    volume = "102",
    number = "8",
    pages = "083024",
    year = "2020"
}

@article{Bayley:2022hkz,
    author = "Bayley, Joseph and Messenger, Chris and Woan, Graham",
    title = "{Rapid parameter estimation for an all-sky continuous gravitational wave search using conditional varitational auto-encoders}",
doi = "10.1103/PhysRevD.106.083022",
    journal = "Phys. Rev. D",
    volume = "106",
    number = "8",
    pages = "083022",
    year = "2022"
}

@article{TaylorApproximants,
    author = "Buonanno, Alessandra and Iyer, Bala and Ochsner, Evan and Pan, Yi and Sathyaprakash, B. S.",
    title = "{Comparison of post-Newtonian templates for compact binary inspiral signals in gravitational-wave detectors}",
doi = "10.1103/PhysRevD.80.084043",
    journal = "Phys. Rev. D",
    volume = "80",
    pages = "084043",
    year = "2009"
}

@book{Maggiore_Vol1,
    author = "Maggiore, Michele",
    title = {{Gravitational Waves: Theory}\\{and Experiments}},
    isbn = "978-0-19-857074-5, 978-0-19-852074-0",
    publisher = "Oxford University Press",
    series = "Oxford Master Series in Physics",
    year = "2007",
    url = {https://global.oup.com/academic/product/gravitational-waves-9780198570745?cc=es&lang=en&},
    pages = {572}
}

@article{abbott2020prospects,
  title={Prospects for observing and localizing gravitational-wave transients with Advanced LIGO, Advanced Virgo and KAGRA},
  author={Abbott, Benjamin P and Abbott, R and Abbott, TD and Abraham, S and Acernese, F and Ackley, K and Adams, C and Adya, VB and Affeldt, C and Agathos, M and others},
  journal={Living reviews in relativity},
  volume={23},
  pages={1--69},
  year={2020},
  publisher={Springer},
 url={http://dx.doi.org/10.1007/s41114-020-00026-9},
}

@article{Miller:2020kmv,
    author = "Miller, Andrew L. and Clesse, S\'ebastien and De Lillo, Federico and Bruno, Giacomo and Depasse, Antoine and Tanasijczuk, Andres",
    title = "{Probing planetary-mass primordial black holes with continuous gravitational waves}",
doi = "10.1016/j.dark.2021.100836",
    journal = "Phys. Dark Univ.",
    volume = "32",
    pages = "100836",
    year = "2021"
}

@article{Miller:2021knj,
    author = "Miller, Andrew L. and Aggarwal, Nancy and Clesse, S\'ebastien and De Lillo, Federico",
    title = "{Constraints on planetary and asteroid-mass primordial black holes from continuous gravitational-wave searches}",
doi = "10.1103/PhysRevD.105.062008",
    journal = "Phys. Rev. D",
    volume = "105",
    number = "6",
    pages = "062008",
    year = "2022"
}

@article{Alestas:2024ubs,
    author = "Alestas, George and Morr\'as, Gonzalo and Yamamoto, Takahiro S. and Garc\'ia-Bellido, Juan and Kuroyanagi, Sachiko and Nesseris, Savvas",
    title = "{Applying the Viterbi algorithm to planetary-mass black hole searches}",
reportNumber = "IFT-UAM/CSIC-24-2",
    doi = "10.1103/PhysRevD.109.123516",
    journal = "Phys. Rev. D",
    volume = "109",
    number = "12",
    pages = "123516",
    year = "2024"
}

@article{article,
    author = {Slade, George},
    title = {The Viterbi algorithm demystified},
    year = {2013},
    month = mar,
    pages = {},
    url={https://api.semanticscholar.org/CorpusID:53639705},
    journal="Semantic Scholar"
    }

@article{10.1093/mnras/224.1.131,
    author = {Schutz, Bernard F. and Tinto, Massimo},
    title = {Antenna patterns of interferometric detectors of gravitational waves – I. Linearly polarized waves},
    journal = {Monthly Notices of the Royal Astronomical Society},
    volume = {224},
    number = {1},
    pages = {131-154},
    year = {1987},
    month = jan,
    issn = {0035-8711},
    doi = {10.1093/mnras/224.1.131},
    url = {https://doi.org/10.1093/mnras/224.1.131}
}

@misc{viterbi_pbh_github,
  title = {{Viterbi tCW Search Pipeline}},
  howpublished = {\url{https://github.com/raulrgdg/viterbi_search-method_PBHs}},
  year = {2026}
}

@article{KAGRA:2023pio,
    author = "Abbott, R. and others",
    collaboration = "KAGRA, VIRGO, LIGO Scientific",
    title = "{Open Data from the Third Observing Run of LIGO, Virgo, KAGRA, and GEO}",
reportNumber = "LIGO-P2200316",
    doi = "10.3847/1538-4365/acdc9f",
    journal = "Astrophys. J. Suppl.",
    volume = "267",
    number = "2",
    pages = "29",
    year = "2023"
}

\end{document}